\pdfoutput=1 
\newif\ifElsevier
\Elseviertrue
\ifElsevier
\documentclass[final,1p]{elsarticle-modified}
\else
\documentclass{journal}
\fi{}

\usepackage[english]{babel}
\usepackage{amssymb,amsfonts,amsmath}
\usepackage {color}
\usepackage[dvipsnames,table]{xcolor}
 \newdimen\footheight 
\usepackage{fullpage}
\usepackage[colorlinks,urlcolor=NavyBlue,citecolor=JungleGreen]{hyperref}

\usepackage{booktabs}
\usepackage{caption}
\usepackage{longtable}

\usepackage{numprint}
\npdecimalsign{.} 

\usepackage{algorithm}
\usepackage{algpseudocode}

\usepackage{pdflscape}
\usepackage{afterpage}

\newcommand{\polylog}{\ensuremath{\mathrm{polylog}}}
\newcommand{\dd}{\ensuremath{\mathrm{deg}}}
\newcommand{\argmin}{\ensuremath{\mathrm{argmin}}}

\ifElsevier
\else

\linenumbers
\usepackage {plaatjes}
\usepackage {a4wide}
\fi{}

\newcommand{\fail}{--}

\newcommand {\etal} {\textit {et al.}}

\renewcommand{\paragraph}[1]{\medskip\noindent\textbf{#1.}}

\newcommand{\alg}[1]{\textbf{#1}}
\newcommand{\dataset}[1]{\texttt{#1}}

\newcommand{\randomheader}{
\centering
\small
\setlength{\tabcolsep}{1.0ex}
\begin{tabular}{rrr@{\hskip 30pt} rr@{\hskip 30pt} rr@{\hskip 30pt} rr@{\hskip 30pt} rr}
\toprule
\multicolumn{3}{c}{Instance}  & \multicolumn{2}{c}{\alg{Tomita}~\cite{tomita-2006}}& \multicolumn{2}{c}{\alg{ELS}~\cite{els-2013}} & \multicolumn{2}{c}{\alg{GreedyBBNX}} & \multicolumn{2}{c}{\alg{GreedyBB}} \\
\cmidrule{1-11}
$n$ & $p$ & $\mu$ & steps & time & steps & time & steps & time & steps & time \\
\midrule
}

\newcommand{\dimacsheader}{
\centering
\small
\setlength{\tabcolsep}{1.0ex}
\begin{tabular}{lrrr@{\hskip 30pt} rr@{\hskip 30pt} rr@{\hskip 30pt} rr@{\hskip 30pt} rr}
\toprule
\multicolumn{4}{c}{Instance}  & \multicolumn{2}{c}{\alg{Tomita}~\cite{tomita-2006}}& \multicolumn{2}{c}{\alg{ELS}~\cite{els-2013}} & \multicolumn{2}{c}{\alg{GreedyBBNX}} & \multicolumn{2}{c}{\alg{GreedyBB}} \\
\cmidrule{1-12}
Name & $d$ & $\omega$ & $\mu$ & steps & time & steps & time & steps & time & steps & time \\
\midrule
}

\newcommand{\newrandomheader}{
\centering
\small
\setlength{\tabcolsep}{1.0ex}
\begin{tabular}{rrr@{\hskip 25pt} rr@{\hskip 25pt} rr@{\hskip 25pt} rr@{\hskip 25pt} rr@{\hskip 25pt} rr@{\hskip 25pt} rr}
\toprule
\multicolumn{3}{c}{Instance}  & \multicolumn{2}{c}{\alg{Naude}~\cite{naude-2016}}& \multicolumn{2}{c}{\alg{TomitaBB}}& \multicolumn{2}{c}{\alg{Tomita}~\cite{tomita-2006}}& \multicolumn{2}{c}{\alg{ELS}~\cite{els-2013}} & \multicolumn{2}{c}{\alg{GreedyBBNX}} & \multicolumn{2}{c}{\alg{GreedyBB}} \\
\cmidrule{1-15}
$n$ & $p$ & $\mu$ & steps & time & steps & time & steps & time & steps & time & steps & time & steps & time \\
\midrule
}

\newcommand{\newdimacsheader}{
\begin{center}
\small
\setlength{\tabcolsep}{1.0ex}
\begin{longtable}{lr@{\hskip 15pt} rr@{\hskip 15pt} rr@{\hskip 15pt} rr@{\hskip 15pt} rr@{\hskip 15pt} rr@{\hskip 15pt} rr}
\toprule
\multicolumn{2}{c}{Instance}  & \multicolumn{2}{c}{\alg{Naude}~\cite{naude-2016}} &  \multicolumn{2}{c}{\alg{TomitaBB}}&\multicolumn{2}{c}{\alg{Tomita}~\cite{tomita-2006}}& \multicolumn{2}{c}{\alg{ELS}~\cite{els-2013}} & \multicolumn{2}{c}{\alg{GreedyBBNX}} & \multicolumn{2}{c}{\alg{GreedyBB}} \\
\cmidrule{1-14}
Name & $\mu$ & steps & time & steps & time & steps & time & steps & time & steps & time & steps & time \\
\midrule
\endhead
\endlastfoot
}

\ifElsevier
\else
\title{Efficiently Listing all Maximal Cliques with Bit-Parallelism}
\shorttitle{Efficiently Listing all Maximal Cliques with Bit-Parallelism}

\author{
  Pablo San Segundo\thanks{Centre for Automation and Robotics (UPM-CSIC),
  Jose Guti\'errez Abascal 2, Madrid 28006, Spain.
  Email: \href{mailto:pablo.sansegundo@upm.es}{pablo.sansegundo@upm.es}
  }
  \and
  Jorge Artieda\footnotemark[1]
  \and
  Darren Strash\thanks{Department of Computer Science, Colgate University, Hamilton, NY, USA.
    Email: \href{mailto:dstrash@cs.colgate.edu}{dstrash@cs.colgate.edu}
  }
}
\shortauthor{San Segundo~\etal}
\fi{}

\begin{document}

\ifElsevier
\begin{frontmatter}
\title{Efficiently Enumerating all Maximal Cliques with Bit-Parallelism}

\author[upm]{Pablo San Segundo\corref{cor1}}
\ead{pablo.sansegundo@upm.es}
\author[upm]{Jorge Artieda}
\author[colgate]{Darren Strash}
\ead{dstrash@cs.colgate.edu}

\cortext[cor1]{Corresponding author}
\address[upm]{Centre for Automation and Robotics (UPM-CSIC),
  Jose Guti\'errez Abascal 2, Madrid 28006, Spain.}
\address[colgate]{Department of Computer Science, Colgate University, Hamilton, NY, USA.}

\begin{abstract}
The maximal clique enumeration (MCE) problem has numerous applications in biology, chemistry, sociology, and graph modeling. Though this problem is well studied, most current research focuses on finding solutions in large sparse graphs or very dense graphs, while sacrificing efficiency on the most difficult medium-density benchmark instances that are representative of data sets often encountered in practice. We show that techniques that have been successfully applied to the maximum clique problem give significant speed gains over the state-of-the-art MCE algorithms on these instances. Specifically, we show that a simple greedy pivot selection based on a fixed maximum-degree first ordering of vertices, when combined with bit-parallelism, performs consistently better than the theoretical worst-case optimal pivoting of the state-of-the-art algorithms of Tomita et al. [Theoretical Computer Science, 2006] and Naud\'e [Theoretical Computer Science, 2016].

Experiments show that our algorithm is faster than the worst-case optimal algorithm of Tomita et al. on 60 out of 74 standard structured and random benchmark instances: we solve 48 instances 1.2 to 2.2 times faster, and solve the remaining 12 instances 3.6 to 47.6 times faster. We also see consistent speed improvements over the algorithm of Naud\'e: solving 61 instances 1.2 to 2.4 times faster. To the best of our knowledge, we are the first to achieve such speed-ups compared to these state-of-the-art algorithms on these standard benchmarks.
\end{abstract}

\end{frontmatter}

\fi{}

\section{Introduction}
\label{section:introduction}
The \emph{maximal clique enumeration} (MCE) problem---the problem of enumerating all maximal cliques of a given graph---has numerous applications spanning many disciplines~\cite{augustson-1970,gardiner99,HorSko-PAMI-89}. Unlike the NP-hard maximum clique problem (MCP)~\cite{Karp1972,garey-johnson-90}, the MCE problem is known to require exponential time in the worst case, since there may be an exponential number of maximal cliques to enumerate. For an $n$-vertex graph, there may be $\Theta(3^{n/3})$ maximal cliques, known as the Moon-Moser bound~\cite{moon-moser-65}, and therefore any algorithm that enumerates all maximal cliques must use at least this amount of time in the worst case. Interestingly, not only does there exist an algorithm that runs in worst-case optimal $\Theta(3^{n/3})$ time, that of Tomita et al.~\cite{tomita-2006}, but it is also among the fastest algorithms in practice. Eppstein et al.~\cite{els-2013} further tightened these bounds for the case of graphs with low \emph{degeneracy}~\cite{LicWhi-CJM-70}, the smallest value $d$ such that every induced subgraph of $G$ has a vertex of degree at most $d$. They showed that graphs with degeneracy $d$ have $\Omega(d(n-d)3^{d/3})$ maximal cliques, and further give an algorithm to enumerate all maximal cliques in time $O(d(n-d)3^{d/3})$, which matches the worst-case output size. Moreover, they showed that their method is efficient in practice on real-world complex networks, which typically have low degeneracy.


These algorithms, as well as many other efficient algorithms, are derived from the Bron-Kerbosch algorithm---which maintains both a currently growing clique and a set of already examined vertices throughout recursive backtracking search, only reporting a clique when it is found to be maximal~\cite{bron-kerbosch-73}. However, a separate class of theoretically-efficient algorithms, those with bounded \emph{time delay} (the time between reported cliques), exist for the MCE problem, which use the reverse search technique of Avis and Fukuda~\cite{avis-1996}. Tsukiyama et al.~\cite{tsukiyama-77} were the first to give a bounded time delay algorithm for this problem, giving an algorithm with delay $O(nm)$ for graphs with $m$ edges. Chiba and Nishizeki~\cite{chiba-nishizeki-1985} improved this result for graphs with arboricity $a$, a sparsity measure, giving a $O(am)$-delay algorithm. Makino and Uno~\cite{makino-uno-2004} removed the linear dependence on $m$, reducing the delay to $O(\Delta^4)$, where $\Delta$ is the maximum degree of $G$; however, their technique uses quadratic space. Chang et al.~\cite{chang-h-index} further showed how to reduce the preprocessing time of Makino and Uno from quadratic to linear, while giving a tighter delay of $O(\Delta h^3)$, where $h$ is the $h$-index of the graph. Finally, Conte et al.~\cite{conte-2016} gave the first bounded delay maximal clique enumeration algorithm with sublinear extra space, with delay $O(qd(\Delta + qd)\polylog(n+m))$, where $q$ is the size of a maximum clique.


Though these bounded time delay algorithms are theoretically efficient, Bron-Kerbosch-derived algorithms are much faster in practice. As noted by Conte et al.~\cite{conte-2016}, their algorithm (which is at present the fastest bounded time delay algorithm) is 3.7 times slower than the Bron-Kerbosch-derived algorithm by Eppstein et al.~\cite{els-2013} on sparse graphs, which is itself slower than the algorithm by Tomita et al.~\cite{tomita-2006} on dense and medium-density instances that we consider here. Even though the algorithm by Tomita et al.~\cite{tomita-2006} has worst-case exponential time, repeated experiments show that it is fast on a variety of benchmark instances~\cite{tomita-2006,els-2013,cazals-karande-2006,koch2001,naude-2016}. At the time of writing, we are unaware of any algorithms that achieve significant speedups over this algorithm, though moderate speedups are possible on graphs that are either very sparse~\cite{els-2013} or very dense~\cite{naude-2016}.


For sparse graphs, the algorithm by Eppstein et al.~\cite{els-2013} rivals that of Tomita et al.~\cite{tomita-2006} while only consuming space linear in the size of the graph, whereas the algorithm of Tomita et al. requires quadratic space to store an adjacency matrix. Dasari et al.~\cite{dasari-2014} further improved this result by factors of 2-4x using bit-parallelism, though the main algorithm remains unchanged. For larger instances, external memory algorithms have been developed which take advantage of the property that real-world sparse graphs typically have small induced subgraphs that can fit into memory~\cite{cheng-2012}. For even larger instances, algorithms have been implemented in the MapReduce framework~\cite{wu-2009}.


In the case of dense graphs, researchers have looked at different strategies for pruning search. In particular, Cazals and Karande~\cite{cazals-karande-2006} showed that detecting and removing dominated vertices is much faster than the traditional pivoting method commonly used in the fastest algorithms, such as that of Tomita et al.~\cite{tomita-2006}, when graphs are dense. This is because the time to pick a pivot can be very expensive when there are many edges. Naud\'e~\cite{naude-2016} investigated the pivot computation set, and showed that it is possible to break out of pivot computation early under certain conditions, and still maintain a worst-case optimal running time of $O(3^{n/3})$. In Naud\'e's experiments on small random graphs (with 180 vertices or less), his algorithm is at most 1.56 times faster than that of Tomita et al.~\cite{tomita-2006} on graphs with a high edge density of 0.8; on the other hand, on graphs with a lower edge density of 0.4, the method gives a modest speed up of at most 13\%.


Surprisingly, recent algorithms have focused only on sparse and high density graphs, and have not considered performance on medium density graphs, which are representative of many real-world instances, and where pruning techniques based on structure, different from the theoretical-optimal pivoting, are much less effective. In particular, the benchmarks from the second DIMACS challenge~\cite{johnson1996} and BHOSLIB~\cite{bhoslib}, have medium density and are among the most difficult sets in practice. As far as we are aware, no algorithm gives significant improvement over the algorithm of Tomita et al.~\cite{tomita-2006} for these instances.


\subsection{Our Contribution}

We provide a new algorithm for the MCE problem, and show that it consistently outperforms the state-of-the-art algorithms of Tomita et al.~\cite{tomita-2006} and Naud\'e~\cite{naude-2016} on benchmark graphs that are representative of difficult instances. 
This work is inspired by a number of techniques that have been described for branch and bound maximum clique solvers, such as employing an initial ordering of nodes~\cite{carraghan-1990,tomita-kameda-2006,tomita-seki-2003,segundo-bitboard-2011,segundo-recoloring,tomita-recoloring}, the use of bit-parallelism~\cite{segundo-bitboard-2011,segundo-recoloring} and keeping a fixed vertex ordering throughout recursion~\cite{segundo-bitboard-2011,segundo-recoloring}. Note that these solvers differ from any MCE solver in the fact that they also prune enumerable cliques in the search tree when they cannot possibly improve the incumbent solution.
%

While leading MCP solvers (as well as the MCE solver by Eppstein et al.~\cite{els-2013}, and the improved bitstring encoding proposed by Dasari et al.~\cite{dasari-2014}) attempt to quickly reduce subproblem size by branching on vertices initially (at the root) with low degree (following a \emph{degeneracy ordering}), we evaluate vertices with high degree according to a \emph{maximum-degree-first ordering}. This ordering helps us address one of the main challenges when applying bit-parallelism to state-of-the-art MCE solvers: efficient \emph{pivot selection}, for which most algorithms must enumerate vertices to find a high degree vertex in the current subproblem~\cite{tomita-2006,els-2013,cazals-karande-2006,koch2001,naude-2016,cazals-karande-tcs}. Moreover, enumeration of items is a well-known bottleneck of bitstrings.

With our proposed initial ordering, we efficiently perform a simple \emph{greedy} pivot selection strategy, which allows us to pivot without enumeration. We likewise branch on vertices according to the initial ordering, since they are likely to have high degree in the subproblem being evaluated. Although this strategy increases the size of the search space on average when compared to the theoretical algorithm of Tomita et al.~\cite{tomita-2006}, our algorithm outperforms state-of-the-art solvers on 60 out of 74 of instances tested, which we attribute to the speed of greedy pivot selection, when combined with a bitstring representation. In contrast, a direct bit-parallel implementation of the state-of-the-art solver by Tomita et al.~\cite{tomita-2006} is slower than the original implementation on most instances.


\subsection{Organization}

The remainder of our paper is organized as follows: in Section~\ref{section:preliminaries}  we cover useful definitions and other preliminaries and in Section~\ref{section:bron-kerbosch}  we describe variations of the Bron-Kerbosch algorithm. Our contributions appear in Section~\ref{section:new-algorithm}, where we describe our new enumeration algorithm. We then present our experimental results in Section~\ref{section:experiments}, and conclude and give ideas for future work in Section~\ref{section:conclusion}.


\section{Preliminaries}
\label{section:preliminaries}
We work with a simple undirected graph $G=(v,E)$, which consists of a finite set of vertices $V=\{1,2,\ldots,n\}$ and a finite set of edges $E\subseteq V \times V$ made up of pairs of distinct vertices. Two vertices $u$ and $v$ are said to be adjacent (or neighbors) if $(u,v)\in E$. The neighborhood of a vertex $v$, denoted $N(v)$ (or $N_G(v)$ when the graph needs to be mentioned explicitly), is defined as $N(v)=\{u\in V\mid(u,v)\in E\}$. We denote the \emph{degree} of a vertex $v$ by $\dd(v)$, and denote the \emph{maximum degree} of $G$ by $\Delta(G) = \max_{v\in V}\dd(v)$.

A \emph{clique}, also called a complete subgraph, is a set $K\subseteq V$ of pairwise adjacent vertices. A clique $K$ is said to be \emph{maximal} if there is no vertex in $V\setminus K$ that is adjacent to all vertices in $K$. Note that this definition is different from a \emph{maximum clique}, which is a clique of maximum cardinality $\omega(G)$.

Finally, we also consider the following terminology and definitions for vertex orderings. We define a vertex ordering to be a sequence $v_1,v_2,\ldots,v_n$ of the vertices in $V$, where $v_i$ is said to be in position $i$, and further define a permutation $\phi:V\rightarrow\{1,2,\ldots,n\}$ that maps each vertex $v_i\in V$ to its position $i$; that is $\phi(v_i) = i$. The \emph{width} of a vertex $v_i$, denoted $w(v_i)$, is the number of vertices adjacent to $v_i$ that precede $v_i$ in a given ordering~\cite{freuder-1982}. We further say that the \emph{width of a vertex ordering} is the maximum width of any of its vertices. The width of a \emph{minimum-width ordering} is also called the \emph{degeneracy} of the graph, which we denote by $d$. An ordering where each vertex has at most $d$ neighbors that come later in the ordering (that is, the \emph{reverse} of a minimum-width ordering) is called a \emph{degeneracy ordering}~\cite{els-2013}. Finally, a degeneracy ordering can be computed in $O(n+m)$ time by iteratively removing a vertex with minimum degree from the graph until it is empty, and placing these vertices in order by their removal~\cite{matula-1983}. A minimum-width ordering can be similarly computed in $O(n+m)$ time by placing vertices in reverse order by their removal.

\section{Existing Bron-Kerbosch Enumeration Algorithms}
\label{section:bron-kerbosch}
In this section, we briefly describe the state-of-the-art techniques for enumerating all cliques in a graph. These algorithms are derived from the Bron-Kerbosch algorithm, which we now describe.

\begin{algorithm}[!htb]
\caption{The Bron-Kerbosch algorithm.}
{\bf proc} BK($P$, $R$, $X$)
\begin{algorithmic}[1]
\If{$P\cup X = \emptyset$}
    \State report $R$ as a maximal clique
\EndIf
\For{ {\bf each} vertex $v\in P$}
    \State BK($P\cap N(v)$, $R\cup\{v\}$, $X\cap N(v)$)
    \State $P \leftarrow P \setminus \{v\}$
    \State $X \leftarrow X \cup \{v\}$
\EndFor
\end{algorithmic}
\label{algorithm:bron-kerbosh}
\end{algorithm}

\subsection{The Bron-Kerbosch Algorithm}
The Bron-Kerbosch algorithm (BK)~\cite{bron-kerbosch-73} is a recursive backtracking algorithm that computes maximal cliques by maintaining three sets of vertices throughout recursion: a clique $R$, a set $P$ of candidates that are to be considered for addition to $R$, and a set $X$ of vertices that have already been evaluated, and thus are excluded from consideration. Throughout execution, BK maintains the invariant that $P\cup X$ is the \emph{common neighborhood} of $R$. That is, $\bigcap_{v\in R}N(v) = P\cup X$. During each recursive call a candidate vertex $v$ from $P$ is moved to $R$, and $P$ and $X$ are updated to maintain this invariant in the next recursive call. Whenever $P$ and $X$ are empty, then $R$ is a maximal clique and it is reported (see Algorithm~\ref{algorithm:bron-kerbosh}). After a vertex $v$ has been evaluated, it is moved to $X$ in step 7. Initially all vertices are candidates (that is, $P=V$), and $R$ and $X$ are empty.

\begin{algorithm}[!htb]
\caption{The Bron-Kerbosch algorithm with pivoting.}
\label{algorithm:bron-kerbosch-pivoting}
{\bf proc} BKPivot($P$, $R$, $X$)
\begin{algorithmic}[1]
\If{$P\cup X = \emptyset$}
    \State report $R$ as a maximal clique
\EndIf
\State $p \leftarrow $ ChoosePivot($P\cup X$)
\For{ {\bf each} vertex $v\in P\setminus N(v)$}
    \State BKPivot($P\cap N(v)$, $R\cup\{v\}$, $X\cap N(v)$)
    \State $P \leftarrow P \setminus \{v\}$
    \State $X \leftarrow X \cup \{v\}$
\EndFor
\end{algorithmic}
\end{algorithm}

\subsection{Bron-Kerbosch with Pivoting}
One technique, which has been highly effective at improving running time, is that of \emph{pivoting}, shown in Algorithm~\ref{algorithm:bron-kerbosch-pivoting}. In pivoting, a vertex $p$ is chosen from either $P$ or $X$, and then only non-neighbors of $p$ in $P$ (including $p$ itself, if $p\in P$) are considered for addition to $R$ in the current recursive call. Such a vertex is called a \emph{pivot}. Koch~\cite{koch2001} initiated a study of different pivoting strategies, including choosing a pivot $p$ from $P$ or $X$ at random or greedily from $P$ to minimize $P\setminus N(p)$ (or equivalently, maximize $|P\cap N(p)|$). From these strategies, Koch showed that selecting from $X$ at random was a superior strategy, closely followed by greedy selection from $P$ (which was slower due to computing the number of neighbors in $P$). However, Cazals and Karande~\cite{cazals-karande-2006} later empirically showed that picking $p$ from $P\cup X$ to minimize $|P\setminus N(p)|$ is even more effective in practice. Tomita et al.~\cite{tomita-2006} showed that this pivoting strategy gives an algorithm with worst-case optimal $O(3^{n/3})$ time.

Naud\'e~\cite{naude-2016} further showed that under certain conditions it is possible to stop pivot selection early, which not only speeds up maximal clique enumeration in practice, but still maintains worst-case optimal running time $O(3^{n/3})$. While Naud\'e's strategy gives a slight running time improvement for graphs with medium density (0.4), speedups of 2x are achieved for graphs with high density (0.8).

\begin{algorithm}[!htb]
\caption{The Bron-Kerbosch algorithm with an initial vertex ordering.}
\label{algorithm:bron-kerbosch-ordering}
{\bf proc} BKOrdering($G=(V,E)$)
\begin{algorithmic}[1]
\State $v_1,v_2,\ldots,v_n \leftarrow $ ComputeOrdering($G$) \Comment{A degeneracy ordering in~\cite{els-2013}}
\For{$i \leftarrow \{1..n\}$}
    \State $P \leftarrow \{v_j\mid j > i\} \cap N(v_i)$
    \State $R \leftarrow \{v_i\}$
    \State $X \leftarrow \{v_j\mid j < i\} \cap N(v_i)$
    \State BKPivot($P$, $R$, $X$)
\EndFor
\end{algorithmic}
\end{algorithm}

\subsection{Bron-Kerbosch with Vertex Ordering}
Vertex ordering, a technique frequently used to solve the maximum clique problem (MCP) can be used to further improve the theoretical running time of BK, while giving fast running times in practice for large sparse graphs. In particular, Eppstein et al.~\cite{els-2013} showed that for graphs with degeneracy $d$, evaluating the vertices in degeneracy order gives running time $O(d(n-d)3^{n/3})$, which matches the worst-case output size.

Previous Bron-Kerbosh-derived algorithms store the input graph in an adjacency matrix, while the method of Eppstein et al.~\cite{els-2013} supports efficient computation using an adjacency list and other linear-sized auxiliary data structures. Thus, their method can be used on large sparse graphs whose adjacency matrix does not fit into memory. By computing an initial ordering of the vertices, they ensure that vertices in $X$ come before vertices of $P$ in the ordering. Note that such a strategy can be generalized to use any initial ordering (see Algorithm~\ref{algorithm:bron-kerbosch-ordering}); however, the key to their algorithm's running time is that the degeneracy ordering ensures that $|P|\leq d$ since each vertex has at most $d$ later neighbors in the ordering.

They further efficiently compute pivots using an auxiliary bipartite graph. Their bipartite graph has vertices $P\cup X$ and $P$ as the left-and right-hand sides respectively, and an edge $(u,v)$ between the two sides when there is an edge in $G$ between $u\in P\cup X$ and $v\in P$. A pivot can be computed by iterating over all neighbor lists in this bipartite graph in time $O(|P|\cdot|P\cup X|)$, and this bipartite graph can be maintained efficiently throughout recursion. For graphs with low degeneracy, their technique rivals the speed of standard pivoting algorithms, while consuming only $O(n+m)$ space. Additionally, their algorithm can be further sped up with bit-parallelism using the strategy of Dasari et al.~\cite{dasari-2014}, giving consistent speed gains of 2-4x.


Lastly, we note that it is possible to apply the same ordering strategy with an adjacency matrix and achieve the same running time. However, it is not clear how this would compare in practice to the previously mentioned algorithms for non-sparse graphs.

\section{The New Enumeration Algorithm}
\label{section:new-algorithm}
Our algorithm combines elements from both Algorithms~\ref{algorithm:bron-kerbosch-pivoting} and~\ref{algorithm:bron-kerbosch-ordering} above. As in Algorithm~\ref{algorithm:bron-kerbosch-ordering}, we compute a vertex ordering. However, unlike Algorithm~\ref{algorithm:bron-kerbosch-ordering}, we perform pivoting at the top level. We further differ with previous maximal clique enumeration algorithms in our set representation, our chosen vertex ordering, and our pivot and vertex selection. See Algorithm~\ref{algorithm:bron-kerbosch-static-ordering} for a high-level overview of the algorithm.

To make effective use of bit-parallelism, we store sets $P$ and $X$ as bitstrings on which we can efficiently perform the necessary set operations. However, to make bit parallelism a viable option, several algorithmic changes are required. Specifically, finding the candidate pivot that minimizes $|P\setminus N(p)|$ is now more expensive than with an array representation of vertex sets, since the vertices in $P$ and $X$ have to be enumerated. We therefore consider a greedy pivot selection strategy based on maintaining an ordering throughout recursion such that `good' pivots (i.e., those that have many neighbors in $P$) appear early in the ordering. Thus, following Algorithm~\ref{algorithm:bron-kerbosch-ordering}, we sort the vertices initially, but instead of computing a degeneracy ordering as in Eppstein et al.~\cite{els-2013} in which each vertex has few later neighbors, we compute an order in which each vertex has many later neighbors---which we call a \emph{maximum-degree-first ordering} (defined formally in Subsection~\ref{subsection:vertex-ordering}). We maintain this relative order of vertices throughout recursion, and use it both for branching and to greedily choose pivots. Note that preserving a static ordering during tree traversal is a significant departure from previous MCE algorithms. We now discuss each component in turn.

\begin{algorithm}[!htb]
\caption{The Bron-Kerbosch algorithm with pivot and vertex selection from the smallest position in the static vertex ordering (given by permutation $\phi$) from those vertices in $X$ or $P$}
\label{algorithm:bron-kerbosch-static-ordering}
{\bf proc} ChoosePivotOrdered($P,X,\phi$)
\begin{algorithmic}[1]
\If{$X \neq \emptyset$}
    \State \Return $\argmin_{v\in X}\phi(v)$
\Else
    \State \Return $\argmin_{v\in P}\phi(v)$
\EndIf
\algstore{bkbreak}
\end{algorithmic}
\quad\\
{\bf proc} BKPivotOrdered($P$, $R$, $X$)
\begin{algorithmic}[1]
\algrestore{bkbreak}
\If{$P\cup X = \emptyset$}
    \State report $R$ as a maximal clique
\EndIf
\State $p \leftarrow $ ChoosePivotOrdered($P$, $X$, $\phi$)
\For{ $v = \argmin_{w\in P\setminus N(p)} \phi(w)$}
    \State BKPivotOrdered($P\cap N(v)$, $R\cup\{v\}$, $X\cap N(v)$, $\phi$)
    \State $P \leftarrow P \setminus \{v\}$
    \State $X \leftarrow X \cup \{v\}$
\EndFor
\end{algorithmic}
\end{algorithm}


\subsection{Bit-parallelism}

\paragraph{Encoding and implementation of critical operations}
Critical sets $P$ and $X$ are encoded as bitstrings, while set $R$ is a classical array since there is no useful bitmask operation which involves $R$. We also use an additional bitset to store the candidate set of vertices $L=P\setminus N(p)$, where $p$ is the pivot selected in each subproblem. The input graph $G$ is also encoded as a list of bitstrings, where each bitset maps to a row of the graph's adjacency matrix, as described in~\cite{segundo-bitboard-2011,segundo-recoloring}. Finally, we also encode the complement graph $\bar{G}$ of $G$ in memory in the same manner. This allows implementing inferences concerned with non-neighbor relations as efficient bitmask operations. 

With the help of the above data structures, the critical operations in Algorithm~\ref{algorithm:bron-kerbosch-static-ordering} are implemented as bitmask AND operations as follows:
\begin{itemize}
\item $P\cap N(v)$ in step 11: AND operation between bitsets $P$ and the $v$-th row of $G$.
\item $X\cap N(v)$ in step 11: AND operation between the $v$-th row of $G$ and bitset $X$.
\item $L=P\setminus N(p)$ in step 10: AND operation between the $p$-th row of $\bar{G}$ and bitset $P$. If the pivot $p$ is in $P$ (that is, not in $X$) then it is further added to the candidate set $L$.
\end{itemize}

Also worth noting is that the empty set test of $P$ and $X$ in step 6 also benefits from the bitset encoding by a constant factor proportional to the number of bitblocks contained in each bitstring.

\subsection{Vertex Ordering}
\label{subsection:vertex-ordering}
We consider an initial \emph{degenerate ordering} (not to be confused with a degeneracy ordering, considered by Eppstein et al.~\cite{els-2013}) which selects at each step, the vertex  with maximum degree---and removes $v$ and all edges incident to $v$---from the original graph. Thus, the resulting order is $v_1,v_2,\ldots,v_n$ where $v_1$ is the vertex with maximum degree in $G$, $v_2$ is the vertex with maximum degree in the graph induced by $V\setminus \{v_1\}$, $v_2$ is the vertex with maximum degree in the subgraph induced by $V\setminus \{v_1, v_2\}$ and so on. The term \emph{degenerate} denotes the fact that the selection criteria are restricted to the remaining vertices and not the full set. We refer to this ordering strategy as a \emph{maximum-degree-first ordering} as opposed to the degeneracy ordering known from literature.


\paragraph{Connection to previous approaches}
Also, in literature, the term \emph{largest-first}~\cite{freuder-1982,welsh-1967}, refers to a (non-degenerate) coloring heuristic which uses an order of vertices based on non-increasing degree; that is, vertices are ordered $v_1,v_2,\ldots,v_n$ such that $\dd(v_1)\geq \dd(v_2)\geq \cdots \geq \dd(v_n)$. The point of this ordering for approximate coloring is to assign color numbers to the most conflicting vertices as early as possible, so that those remaining will require a small number of colors to complete the partial coloring. 

A branch-and-bound algorithm for the exact MCP is usually concerned with two vertex orderings, one for branching initially and the other for approximate-coloring---a critical part of the so-called \emph{bounding function}, since the number of distinct colors required to color a graph $G$ is an upper bound for the cardinality $\omega(G)$ of a maximum clique in the graph. In the case of BBMC~\cite{segundo-bitboard-2011,segundo-recoloring} and MCS~\cite{tomita-recoloring}, two leading algorithms for the MCP, branching at the root node follows a degeneracy ordering where the vertex ordering $v_1,v_2,\ldots,v_n$ is produced by iteratively removing a vertex with minimum degree from $G$ along with its incident edges, and $v_i$ is the $i$-th vertex removed in this way.

This strategy is well known to reduce the size of the search tree, in some cases even exponentially, by attempting to minimize branching in the shallower levels of the search tree (vertices with low degrees, produce small subproblems with high probability). In the remaining subproblems, both algorithms then switch to branching on a maximum-color-label basis; but the relative order of vertices remains the same (that is, as determined initially) in all subproblems. Consequently, the initial order also implicitly conditions the approximate-coloring bounding procedure.

Although no such bounding function exists for the MCE, we follow a similar approach and also preserve the initial order of vertices in all subproblems. We note that, although Eppstein et al.~\cite{els-2013} described an ordering heuristic for the MCE, their ordering is only applied at the beginning of the algorithm, and is not maintained throughout recursion. To the best of our knowledge, we are the first to explore this strategy for the MCE.


\paragraph{How we use the vertex ordering}
We briefly mention two implementation details that involve our initial vertex ordering.

\begin{itemize}
\item Once we compute the \emph{maximum-degree-first} ordering, we place the vertices in \emph{reverse order} (and branch on vertices in reverse order as well, from last to first). Note that this is a practical consideration consistent with previous algorithms for MCP, such as MCS or BBMC. 
\item We renumber each vertex according to its position in the initial ordering from left to right, and reconstruct the adjacency matrix with respect to this numbering, i.e. we build the graph isomorphism which corresponds to the new ordering. This was done by Tomita et al.~\cite{tomita-recoloring} for the MCS algorithm, and for the bit-parallel algorithm BBMC by San Segundo et al.~\cite{segundo-bitboard-2011,segundo-recoloring,segundo-implicit}. Both authors note that renumbering ensures locality of data. The new adjacency matrix sets the reference for all bitsets; consequently, the relative order of vertices is preserved in (bitsets) $P$ and $X$ in every subproblem.
\end{itemize}

By maintaining the initial vertex ordering throughout recursion, we are able to quickly select a vertex with many neighbors in the graph, which is likely to have a high degree in the current subproblem. We use this fact to efficiently (and greedily) compute a pivot vertex, which we now describe.





\subsection{Pivot Selection}
As explained previously, we use a \emph{maximum-degree-first} initial ordering for our new algorithm (and maintain it in all subproblems). 

The key idea behind this strategy is to use the initial vertex ordering as an implicit pivoting heuristic to quickly find pivots with many neighbors in $P$, and thus reduce the branching factor. Initially, vertices are in \emph{maximum-degree-first} order and the first vertex $p$ in the ordering is chosen both as pivot and as starting point of the search. This is consistent with the theoretical optimal pivoting in~\cite{tomita-2006} since it maximizes $|P\cap N(p)|$ and, consequently, minimizes branching in step 10 of Algorithm~\ref{algorithm:bron-kerbosch-static-ordering}. From then on, we greedily select each new pivot from candidate pivot set $P\cup X$, choosing the vertex $p\in X$ (or is $X$ is empty, the vertex in $P$) that appears earliest in the initial ordering among all vertices in $X$ (or $P$), which we call the first vertex of $X$ (or $P$). We further always branch (in step 10) on the first vertex of $P$. Since each new pivot is has many later neighbors in the ordering, the hope is that it will also have many neighbors in $P$, thereby increasing the likelihood of maximizing $|P\cap N(p)|$ in future subproblems. Prioritizing set $X$ with respect to $P$ in our greedy pivot selection is also intended to reduce $|P\setminus N(p)|$, since any pivot $p$ in $X$ cannot itself make part of the branching set. That is, $p\not\in P\setminus N(p)$ if $p\in X$.

Notice that this greedy selection is accomplished in constant time, since we maintain our maximum-degree-first ordering throughout recursion. Thus, the running time for our pivot selection is much faster than other pivot selection routines. For example, the pivoting strategy by Tomita et al.~\cite{tomita-2006} iterates over all vertices and counts their neighbors in $P$, taking $\Theta(n^2)$ time in the worst case. Unlike Tomita et al., Naud\'e's~\cite{naude-2016} pivoting strategy also adds vertices to $P$, increasing the time to select a pivot vertex. His strategy takes $\Theta((k+1)n^2)$ time, where $k$ is the number of vertices added to $P$, and thus takes $\Theta(n^3)$ time in the worst case.

Note that the algorithms of Tomita et al.~\cite{tomita-2006} and Naud\'e~\cite{naude-2016} are worst-case optimal, and their analysis depends on the fact that \emph{some} elements from $P$ are excluded from immediate recursive calls. In contrast, our greedy pivoting strategy provides no such guarantee. Therefore, it is unclear if our algorithm is any more efficient in the worst case than the standard Bron-Kerbosch algorithm without pivoting, for which no non-trivial analysis is known.

Notice also that our greedy selection strategy differs from the greedy pivot strategies of Koch~\cite{koch2001}, Cazals and Karande~\cite{cazals-karande-2006}, and Tomita et al.~\cite{tomita-2006}. They select a vertex $p$ maximizing $|P\cap N(p)|$ whereas we select a vertex that has many neighbors in  (which hopefully, by extension, has many neighbors in $P$), since this selection is fast according to our static ordering.

\paragraph{Different pivot strategies}
Similar to previous approaches~\cite{tomita-2006,cazals-karande-2006,koch2001,johnston75}, our greedy pivot selection can be applied to the full pivot candidate set $P\cup X$, or selectively to sets $P$ and $X$. Experiments by Koch~\cite{koch2001} and Cazals and Karande~\cite{cazals-karande-2006} show that selecting pivots from $X$ (and in Cazals and Karande's case, considering $X$ in addition to $P$) is stronger than selecting pivots from $P$ alone. In all of our algorithms, we therefore first select a pivot from $X$, and only consider pivot candidates from $P$ if $X$ is empty. This method is similar to Koch's variant that selects random pivots exclusively from $X$~\cite{koch2001}; however, it is not clear if Koch's variant also selects a pivot from $P$ when $X$ is empty. Notwithstanding, we differ in our pivot selection criteria: we select $p$ with many neighbors in $G$, attempting to maximize the number of neighbors in $P$ without explicitly computing the optimal pivot.

We consider the following strategies that take advantage of this greedy technique. In all of them, ties are broken in favor of the index with smaller index (in practice, larger since, owing to practical considerations, we order vertices by maximum degree in \emph{reverse} order).

\alg{GreedyBB}. In this algorithm, we select as pivot the vertex $p$ that is the first vertex in $X$. If $X$ is empty, then the pivot becomes the first vertex in $P$. This is our main variant. As mentioned earlier, the idea of selecting pivots from $X$ (if $X$ is not empty) instead of from $P\cup X$ is based on the fact that any pivot in $X$ will not be in $P$, so this reduces by one (the pivot) the branching factor at every step. 


We further consider two variants that use more informed strategies to select a pivot vertex from $X$:
\begin{itemize}
\item \alg{GreedyBBTX}. Here pivot selection is as follows: if $X$ is not empty, then the pivot $p$ is a vertex from $X$ which maximizes $|P\cap N(p)|$ (as in the algorithm of Tomita et al.~\cite{tomita-2006}). If $X$ is empty, $p$ is the first vertex in $P$.
\item \alg{GreedyBBNX}. This algorithm is the same as \alg{GreedyBBTX}, however, it includes the optimizations introduced by Naud\'e~\cite{naude-2016} when computing pivot $p$ from vertices in $X$. Naud\'e considers the mathematical equivalent expression of minimizing $|P\cap K(p)|$, where $K(p)=V\setminus N(p)$ (including $p$ itself). The advantage of this formulation is that it is possible to stop evaluating a candidate pivot $p$ when it cannot improve the best pivot found so far.
\end{itemize}

We compare the new algorithm with the original algorithm by Tomita et al.~\cite{tomita-2006} (\alg{Tomita}), provided to us by its main developer, and the critical optimizations by Naud\'e~\cite{naude-2016} (\alg{Naude}). Moreover, we also compare our own bit-parallel implementation of \alg{Tomita} (\alg{TomitaBB}). \alg{TomitaBB} differs from the original algorithm in the bitstring encoding. It also sorts vertices initially as in \alg{GreedyBB} but, since pivots are selected in the theoretically optimal way, this should not alter performance significantly.


\section{An Example}

\begin{figure}[!tb]
\centering
\includegraphics{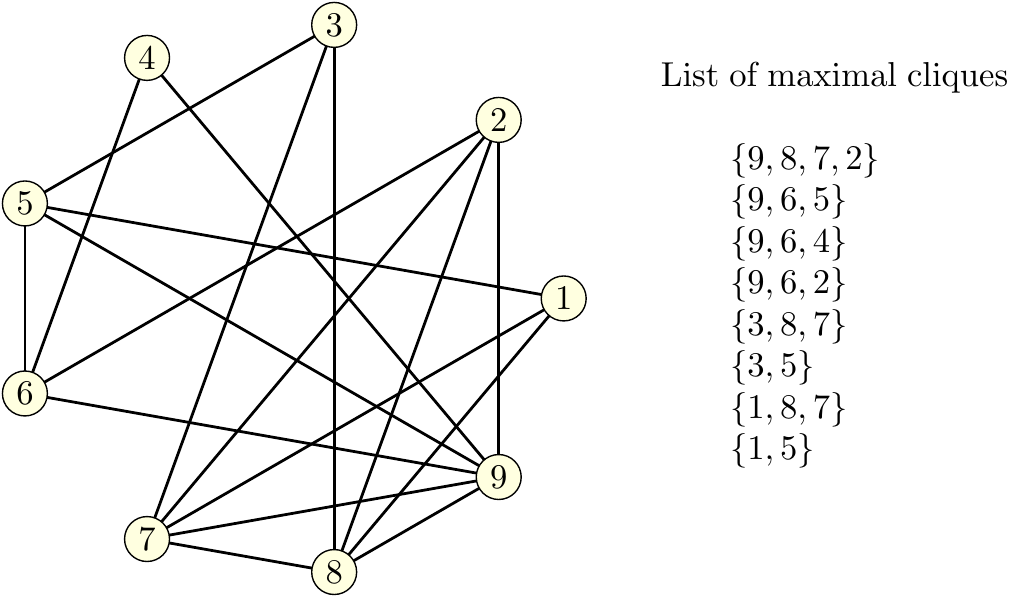}
\caption{A demonstration graph ordered with maximum-degree-first order $9, 8, 7, 6, 5, 4, 3, 2, 1$: vertex $9$ has maximum degree in $V$, vertex $8$ has maximum degree in $V\setminus\{9\}$, vertex $7$ has maximum degree in $V\setminus\{9,8\}$, and so on. The order is reversed in accordance with the current implementation.}
\label{figure:example}
\end{figure}

Since our greedy pivot selection strategy is simpler than the theoretically optimal one~\cite{tomita-2006}, it is to be expected that the new algorithm examines more subproblems. However, the simplicity of our greedy strategy makes pivot selection much faster. We would therefore expect our algorithm to outperform the optimal one if this speed is enough to offset the time spent in the additional subproblems.

We justify the ``good'' behavior of the new greedy pivot selection strategy with the help of the graph $G$ depicted in Figure~\ref{figure:example}. The graph is sorted initially according to maximum-degree-first and the list of the $8$ maximal cliques to be found appear on the right. We compare the behavior of \alg{GreedyBB} and \alg{TomitaBB} for this case. As noted, both algorithms will always select vertices with highest degree first for branching---at the root node, vertex $9$. The new set of candidates $P^\prime = P\cap N(9)$ is $\{8(2),7(2),6(3),5(1),4(1),2(3)\}$, where the parentheses contain the degree of each vertex in $P^\prime$.

\alg{GreedyBB} now selects as new pivot $8$, the vertex with highest degree in $P^\prime$, since the conflict set is still empty. Consequently, the set of vertices $L^\prime_{GBB}=P^\prime\setminus N(8)$ to be expanded in $P^\prime$ (step 5 of Algorithm~\ref{algorithm:bron-kerbosch-pivoting}) becomes $\{8,6,5,4\}$. On the other hand, \alg{TomitaBB} selects as pivot vertex $6$ because it has maximum degree in $P^\prime$. This reduces the set of vertices to be expanded in $P^\prime$ to $L^\prime_{TBB}=\{8,7,6\}$.

Although $|L^\prime_{TBB}|<|L^\prime_{GBB}|$, it turns out that the choice of pivot made by \alg{TomitaBB} is suboptimal. This can be established by looking at the listing of maximal cliques in the figure. \alg{GreedyBB} first takes $8$ from $L^\prime_{GBB}$---to find the maximum clique $\{9,8,7,2\}$---and then $6$, to enumerate the family of cliques $\{9,6,*\}$. This is optimal and the search requires a total of $8$ steps, where a step is a recursive call to the algorithm. \alg{TomitaBB}, however, requires $9$ steps, because it branches suboptimally on vertex $7$ (instead of $6$) after backtracking from $8$. This leads to the clique $\{9,7,2\}$, which is not maximal.

To summarize, the new greedy pivot selection strategy defeats the standard strategy in the example because the neighbor set of the \alg{TomitaBB} pivot $6$ \emph{contains only vertices preceding the pivot}. It does not contain vertex $7$, which has higher index and, therefore, higher probability of belonging to a large maximal clique---in the example, the maximum clique.

\section{Experiments}
\label{section:experiments}
\paragraph{Setup}
Experiments were performed using a Linux workstation running a 64-bit Ubuntu release at 3.00 GHz with an Intel(R) Xeon(R) CPU E5-2690 v2 multi-core processor (two sets of 10 physical cores) and 128 GB of main memory. All algorithms tested were implemented in C or C++, compiled using \texttt{gcc} version 4.8.1 with the \texttt{-O3} optimization flag, and each algorithm was run on a dedicated core.

\paragraph{Data sets}
We run the algorithms over a number of structured and uniform random graphs. Specifically, structured graphs are those from the well-known DIMACS clique and color data sets (from the second DIMACS implementation challenge~\cite{johnson1996}). The DIMACS clique graphs selected include those typically employed elsewhere, as well as new ones which were computed in less than 6 hours by the new algorithm. In the case of the less dense color data set, we report those that required the Tomita algorithm more than 0.2 seconds to solve. Moreover, we also considered the BHOSLIB benchmark~\cite{bhoslib}, but none of the graphs tested run inside the 6-hour time limit. With respect to uniform random graphs, the concrete set of instances shown in the report is borrowed from Eppstein et al.~\cite{els-2013}.

\paragraph{Algorithms tested}
In our tests, we consider the bit-parallel implementations of the new greedy pivot strategy: \alg{GreedyBB}, \alg{GreedyBBTX}, and \alg{GreedyBBNX}, as well our implementation of the bit-parallel of the algorithm by Tomita et al.~\cite{tomita-2006} \alg{TomitaBB}. We further test the original algorithms by Tomita et al.~\cite{tomita-2006}, Naud\'e~\cite{naude-2016} and Eppstein et al.~\cite{els-2013}. We use the source code as provided by the main developers---the latter is the same code used in the experiments of Eppstein et al.~\cite{els-2013}\footnote{The source code is available at \url{https://github.com/darrenstrash/quick-cliques/releases}.}. The source code for~\cite{naude-2016} has also been recently published\footnote{The source code is available at \url{https://github.com/kevin-a-naude/cliques}.}, and some modifications were made on demand by the developer to enable us to compute the number of steps (nodes in the search space). We now briefly describe each of these algorithms:



\begin{itemize}
\item \alg{Tomita}: This is the original adjacency matrix implementation of the algorithm by Tomita et al.~\cite{tomita-2006}.
\item \alg{Naude}: The original implementation of the optimizations over \alg{Tomita} described in~\cite{naude-2016}. We note that there is no fundamental change in the theoretical pivoting of the former.
\item \alg{ELS}: This is the algorithm by Eppstein et al.~\cite{els-2013} computes a degeneracy ordering, evaluates vertices in degeneracy order (as discussed in Section~\ref{section:bron-kerbosch}), and quickly computes a pivot in $O(|P|\cdot|P\cup X|)$ time by dynamically maintaining a bipartite graph between vertices in $P\cup X$ and $P$. We briefly note that the \alg{ELS} algorithm is tailored to work efficiently on sparse graphs. However, we include it in our experiments since it is the only existing MCE algorithm that uses an ordering strategy.
\end{itemize}

\subsection{Experimental Results}
We now compare the performance of the three prior leading algorithms, \alg{Tomita}, \alg{Naude} and \alg{ELS}, with the two best new bit-parallel algorithms: our main variant \alg{GreedyBB}, \alg{GreedyBBNX} and our bit-parallel implementation of~\cite{tomita-2006}, \alg{TomitaBB}, (see Table~\ref{table:random} and Table~\ref{table:dimacs}). The full results obtained by all the algorithms considered in this research are publicly available at~\cite{enum-results}. The source code may be found at~\cite{enum-code}.

The best performance is achieved undoubtedly by \alg{GreedyBB}, which uses pure greedy pivot selection. Of the 28 different uniformly random graphs considered in Table~\ref{table:random}, it is faster than \alg{ELS} in 25 cases---and \alg{GreedyBB} is only slower than \alg{ELS} on the sparse graphs with \numprint{10000} nodes, for which \alg{ELS} was designed---faster than \alg{Tomita} in all but 4 cases, and always faster than \alg{GreedyBBNX}. Moreover, in the 46 structured graphs reported in Table~\ref{table:dimacs}, \alg{GreedyBB} is faster than \alg{Tomita} in 36 cases, faster than \alg{GreedyBBNX} in 40, and faster than \alg{Naude} in 38. Moreover, it is the only algorithm, together with \alg{Naude} that solves the hard instance \dataset{p\_hat500.2} within the 6-hour time limit.

Furthermore, the fast speed of our algorithm is only partially due to bit-parallelism. Using bit-parallelism, we can expect an algorithm's running time to decrease by a constant factor $cW$ where is the size of the register word (in our case $W=64$) and $c<1$ is a constant which models the cost of inefficient operations such us bit-scanning loops. Bit-scanning is required by the theoretical pivot selection strategy, since it requires enumeration of candidates. Thus, it is the \emph{combined new greedy pivot selection with the bitstring representation} that explains the practical value of \alg{GreedyBB}. We can see this in the performance of \alg{TomitaBB}. In the 46 structured instances, \alg{GreedyBB} is better in most cases, and never worse. Also, in the larger uniform random graphs ($n>1000$) the performance of \alg{TomitaBB} deteriorates considerably because the enumeration during pivoting is now over large bitstrings.

It is worth noting at this point, that bitstrings are particularly well suited to preserve the order of elements when encoding set operations. Because of this, pivoting based on relative order is very fast in \alg{GreedyBB}. This explains why it clearly outperforms not only \alg{TomitaBB}, but also the original \alg{Tomita} in the large non-structured graphs. 

\alg{GreedyBB} achieves speedups ranging from 1.2 to 20 times \alg{Tomita} in the structured graphs, but, in the majority of these cases, does not double the speed of the latter. In those graphs where it is more than 3 times faster, \dataset{san400\_0.5\_1} constitutes a good example of the success of the greedy pivoting strategy over a difficult instance. In the case of uniform random graphs, the improvement in performance of \alg{GreedyBB} reaches up to 47 times, and is especially relevant in the large graphs as mentioned previously.

If we look at the number of recursive calls (also steps) taken by the algorithms, \alg{GreedyBB} requires more steps than the theoretical algorithm, on average, when the graphs have some structure. This was to be expected. Notable exceptions are \dataset{dsjc500.1}, \dataset{dsjc1000.1}, \dataset{hamming6-4}, \dataset{johnson16-2-4}, \dataset{johnson8-2-4} and \dataset{abb313GPIA}. However, the cases where the number of calls doubles the theoretical algorithm are few, and degenerate behavior is only observed in the 2 graphs reported from the \emph{wap} family. Interestingly, \alg{GreedyBB} is only a 25\% slower than \alg{Tomita} over \emph{wap}, which illustrates that the synergies between the new pivoting rule and the bistring encoding achieve a good compromise between pruning and computational overhead.

In the case of uniform random graphs, \alg{GreedyBB} \emph{makes fewer calls than the theoretical algorithm in 14 cases out of the 28 families reported}. This, we believe to be an interesting, and unexpected, result. The majority of cases in which the greedy pivoting strategy produces a smaller search tree occur in the less dense graphs. As density increases, the theoretical pivoting selection gradually prunes better, which is consistent with intuition.

\section{Conclusion and Future Work}
\label{section:conclusion}
We have shown that a greedy pivot selection strategy is consistently 1.2 to 2.4 times faster than the state-of-the-art algorithms of Tomita et al.~\cite{tomita-2006} and Naud\'e~\cite{naude-2016} for many structured and unstructured benchmark instances when combined with bit-parallelism and a static maximum-degree-first ordering of the vertices. Though similar techniques have long been known to be successful in solving the MCP, this is the first time, to the best of our knowledge, they are being applied to MCE. Moreover, ordering vertices by maximum degree is \emph{not} a standard ordering used by exact MCP solvers, as it has been observed to slow down search~\cite{tomita-kameda-2006}. In contrast, they branch on those vertices with smaller degree at the root node, which tends to produce small subproblems in the shallower levels of the tree.

The success of our approach leaves open three major questions for future research. First, are there other techniques from MCP algorithms that can further improve algorithms for MCE? Next, are there other simple pivot selection methods that can further improve MCE algorithms? Finally, is it possible to gain more speed from bit-parallelism when combined with pivoting in MCE algorithms?



\section*{Acknowledgments}
This work is funded by the Spanish Ministry of Economy and Competitiveness (grant NAVEGASE: DPI 2014-53525-C3-1-R).

\section*{References}

\bibliographystyle{elsarticle-num} 
\bibliography{bitcliques}

\clearpage
\pagestyle{empty}

\afterpage{%
\clearpage
\begin{landscape}
\begin{table}[htb]
\caption{Performance of different clique enumeration algorithms over uniform random graphs of size $n$ and uniform density $p$. Cells show average values over 10 runs. Column header $\mu$ is the number of maximal cliques found. In bold, the best performance for each row. Times are measured in seconds with millisecond precision.}
\label{table:random}
\footnotesize
\centering
\newrandomheader
\numprint{100} & \numprint{0.6} & \nprounddigits{1}\numprint{61.658}K\npnoround & \nprounddigits{1}\numprint{122.347}K\npnoround & \numprint{0.019} & \nprounddigits{1}\numprint{85.480}K\npnoround & \numprint{0.022} & \nprounddigits{1}\numprint{90.441}K\npnoround & \numprint{0.019} & \nprounddigits{1}\numprint{153.163}K\npnoround & \numprint{0.075} & \nprounddigits{1}\numprint{113.980}K\npnoround & \numprint{0.016} & \nprounddigits{1}\numprint{144.778}K\npnoround & \textbf{\numprint{0.009}} \\
\numprint{100} & \numprint{0.7} & \nprounddigits{1}\numprint{407.993}K\npnoround & \nprounddigits{1}\numprint{806.976}K\npnoround & \numprint{0.108} & \nprounddigits{1}\numprint{575.968}K\npnoround & \numprint{0.140} & \nprounddigits{1}\numprint{604.355}K\npnoround & \numprint{0.122} & \nprounddigits{1}\numprint{1.015376}M\npnoround & \numprint{0.412} & \nprounddigits{1}\numprint{850.090}K\npnoround & \numprint{0.089} & \nprounddigits{1}\numprint{1.074488}M\npnoround & \textbf{\numprint{0.061}} \\
\numprint{100} & \numprint{0.8} & \nprounddigits{1}\numprint{5.765723}M\npnoround & \nprounddigits{1}\numprint{11.233850}M\npnoround & \numprint{1.175} & \nprounddigits{1}\numprint{7.998545}M\npnoround & \numprint{1.526} & \nprounddigits{1}\numprint{8.320315}M\npnoround & \numprint{1.532} & \nprounddigits{1}\numprint{14.122934}M\npnoround & \numprint{5.026} & \nprounddigits{1}\numprint{13.458355}M\npnoround & \numprint{0.983} & \nprounddigits{1}\numprint{15.349167}M\npnoround & \textbf{\numprint{0.827}} \\
\numprint{100} & \numprint{0.9} & \nprounddigits{1}\numprint{293.383932}M\npnoround & \nprounddigits{1}\numprint{570.384721}M\npnoround & \numprint{51.900} & \nprounddigits{1}\numprint{401.376665}M\npnoround & \numprint{70.235} & \nprounddigits{1}\numprint{412.908783}M\npnoround & \numprint{63.986} & \nprounddigits{1}\numprint{706.838762}M\npnoround & \numprint{214.008} & \nprounddigits{1}\numprint{751.542684}M\npnoround & \numprint{47.112} & \nprounddigits{1}\numprint{762.601374}M\npnoround & \textbf{\numprint{39.479}} \\ \hline
\numprint{300} & \numprint{0.1} & \nprounddigits{1}\numprint{3.773}K\npnoround & \nprounddigits{1}\numprint{6.215}K\npnoround & \numprint{0.002} & \nprounddigits{1}\numprint{2.516}K\npnoround & \numprint{0.004} & \nprounddigits{1}\numprint{3.611}K\npnoround & \textbf{$<$\numprint{0.001}} & \nprounddigits{1}\numprint{7.368}K\npnoround & \numprint{0.007} & \nprounddigits{1}\numprint{2.589}K\npnoround & \numprint{0.002} & \nprounddigits{1}\numprint{2.621}K\npnoround & \numprint{0.001} \\
\numprint{300} & \numprint{0.2} & \nprounddigits{1}\numprint{18.360}K\npnoround & \nprounddigits{1}\numprint{34.085}K\npnoround & \numprint{0.007} & \nprounddigits{1}\numprint{17.151}K\npnoround & \numprint{0.012} & \nprounddigits{1}\numprint{22.154}K\npnoround & \textbf{$<$\numprint{0.001}} & \nprounddigits{1}\numprint{40.365}K\npnoround & \numprint{0.024} & \nprounddigits{1}\numprint{18.512}K\npnoround & \numprint{0.010} & \nprounddigits{1}\numprint{19.976}K\npnoround & \numprint{0.006} \\
\numprint{300} & \numprint{0.3} & \nprounddigits{1}\numprint{91.395}K\npnoround & \nprounddigits{1}\numprint{183.931}K\npnoround & \numprint{0.040} & \nprounddigits{1}\numprint{107.198}K\npnoround & \numprint{0.062} & \nprounddigits{1}\numprint{125.907}K\npnoround & \numprint{0.029} & \nprounddigits{1}\numprint{216.818}K\npnoround & \numprint{0.122} & \nprounddigits{1}\numprint{119.690}K\npnoround & \numprint{0.031} & \nprounddigits{1}\numprint{140.650}K\npnoround & \textbf{\numprint{0.023}} \\
\numprint{300} & \numprint{0.4} & \nprounddigits{1}\numprint{526.796}K\npnoround & \nprounddigits{1}\numprint{1.124871}M\npnoround & \numprint{0.235} & \nprounddigits{1}\numprint{729.327}K\npnoround & \numprint{0.341} & \nprounddigits{1}\numprint{813.507}K\npnoround & \numprint{0.210} & \nprounddigits{1}\numprint{1.334771}M\npnoround & \numprint{0.752} & \nprounddigits{1}\numprint{844.271}K\npnoround & \numprint{0.181} & \nprounddigits{1}\numprint{1.116905}M\npnoround & \textbf{\numprint{0.125}} \\
\numprint{300} & \numprint{0.5} & \nprounddigits{1}\numprint{4.350726}M\npnoround & \nprounddigits{1}\numprint{9.681350}M\npnoround & \numprint{1.807} & \nprounddigits{1}\numprint{6.799166}M\npnoround & \numprint{3.097} & \nprounddigits{1}\numprint{7.288962}M\npnoround & \numprint{1.888} & \nprounddigits{1}\numprint{11.582656}M\npnoround & \numprint{6.501} & \nprounddigits{1}\numprint{8.275187}M\npnoround & \numprint{1.628} & \nprounddigits{1}\numprint{12.635749}M\npnoround & \textbf{\numprint{1.120}} \\
\numprint{300} & \numprint{0.6} & \nprounddigits{1}\numprint{64.781212}M\npnoround & \nprounddigits{1}\numprint{146.664453}M\npnoround & \numprint{26.557} & \nprounddigits{1}\numprint{108.561034}M\npnoround & \numprint{48.960} & \nprounddigits{1}\numprint{113.689289}M\npnoround & \numprint{29.062} & \nprounddigits{1}\numprint{177.280125}M\npnoround & \numprint{95.941} & \nprounddigits{1}\numprint{141.525597}M\npnoround & \numprint{25.258} & \nprounddigits{1}\numprint{252.730263}M\npnoround & \textbf{\numprint{20.336}} \\ \hline
\numprint{500} & \numprint{0.1} & \nprounddigits{1}\numprint{15.046}K\npnoround & \nprounddigits{1}\numprint{24.191}K\npnoround & \numprint{0.005} & \nprounddigits{1}\numprint{9.470}K\npnoround & \numprint{0.014} & \nprounddigits{1}\numprint{12.464}K\npnoround & \textbf{$<$\numprint{0.001}} & \nprounddigits{1}\numprint{27.447}K\npnoround & \numprint{0.016} & \nprounddigits{1}\numprint{9.799}K\npnoround & \numprint{0.009} & \nprounddigits{1}\numprint{10.001}K\npnoround & \numprint{0.004} \\
\numprint{500} & \numprint{0.2} & \nprounddigits{1}\numprint{100.261}K\npnoround & \nprounddigits{1}\numprint{187.954}K\npnoround & \numprint{0.042} & \nprounddigits{1}\numprint{96.014}K\npnoround & \numprint{0.070} & \nprounddigits{1}\numprint{121.023}K\npnoround & \numprint{0.031} & \nprounddigits{1}\numprint{220.277}K\npnoround & \numprint{0.128} & \nprounddigits{1}\numprint{102.964}K\npnoround & \numprint{0.038} & \nprounddigits{1}\numprint{113.128}K\npnoround & \textbf{\numprint{0.023}} \\
\numprint{500} & \numprint{0.3} & \nprounddigits{1}\numprint{711.562}K\npnoround & \nprounddigits{1}\numprint{1.484998}M\npnoround & \numprint{0.343} & \nprounddigits{1}\numprint{883.318}K\npnoround & \numprint{0.610} & \nprounddigits{1}\numprint{1.038695}M\npnoround & \numprint{0.278} & \nprounddigits{1}\numprint{1.742923}M\npnoround & \numprint{1.093} & \nprounddigits{1}\numprint{979.684}K\npnoround & \numprint{0.302} & \nprounddigits{1}\numprint{1.198122}M\npnoround & \textbf{\numprint{0.157}} \\
\numprint{500} & \numprint{0.4} & \nprounddigits{1}\numprint{6.503028}M\npnoround & \nprounddigits{1}\numprint{14.688254}M\npnoround & \numprint{3.428} & \nprounddigits{1}\numprint{9.730605}M\npnoround & \numprint{6.701} & \nprounddigits{1}\numprint{10.817653}M\npnoround & \numprint{2.992} & \nprounddigits{1}\numprint{17.218052}M\npnoround & \numprint{11.238} & \nprounddigits{1}\numprint{11.178338}M\npnoround & \numprint{3.221} & \nprounddigits{1}\numprint{15.838666}M\npnoround & \textbf{\numprint{1.809}} \\
\numprint{500} & \numprint{0.5} & \nprounddigits{1}\numprint{96.633609}M\npnoround & \nprounddigits{1}\numprint{231.402732}M\npnoround & \numprint{53.703} & \nprounddigits{1}\numprint{164.783219}M\npnoround & \numprint{113.829} & \nprounddigits{1}\numprint{177.168722}M\npnoround & \numprint{49.815} & \nprounddigits{1}\numprint{271.155752}M\npnoround & \numprint{172.629} & \nprounddigits{1}\numprint{197.765682}M\npnoround & \numprint{53.718} & \nprounddigits{1}\numprint{338.113334}M\npnoround & \textbf{\numprint{35.069}} \\ \hline
\numprint{700} & \numprint{0.1} & \nprounddigits{1}\numprint{37.702}K\npnoround & \nprounddigits{1}\numprint{61.599}K\npnoround & \numprint{0.014} & \nprounddigits{1}\numprint{23.638}K\npnoround & \numprint{0.028} & \nprounddigits{1}\numprint{31.183}K\npnoround & \textbf{\numprint{0.010}} & \nprounddigits{1}\numprint{68.638}K\npnoround & \numprint{0.033} & \nprounddigits{1}\numprint{24.731}K\npnoround & \numprint{0.014} & \nprounddigits{1}\numprint{25.324}K\npnoround & \numprint{0.011} \\
\numprint{700} & \numprint{0.2} & \nprounddigits{1}\numprint{327.804}K\npnoround & \nprounddigits{1}\numprint{630.377}K\npnoround & \numprint{0.163} & \nprounddigits{1}\numprint{321.344}K\npnoround & \numprint{0.310} & \nprounddigits{1}\numprint{400.682}K\npnoround & \numprint{0.121} & \nprounddigits{1}\numprint{725.419}K\npnoround & \numprint{0.461} & \nprounddigits{1}\numprint{345.019}K\npnoround & \numprint{0.145} & \nprounddigits{1}\numprint{382.756}K\npnoround & \textbf{\numprint{0.076}} \\
\numprint{700} & \numprint{0.3} & \nprounddigits{1}\numprint{3.058419}M\npnoround & \nprounddigits{1}\numprint{6.600125}M\npnoround & \numprint{1.785} & \nprounddigits{1}\numprint{3.959827}M\npnoround & \numprint{3.772} & \nprounddigits{1}\numprint{4.646231}M\npnoround & \numprint{1.332} & \nprounddigits{1}\numprint{7.663545}M\npnoround & \numprint{5.451} & \nprounddigits{1}\numprint{4.377145}M\npnoround & \numprint{1.705} & \nprounddigits{1}\numprint{5.482722}M\npnoround & \textbf{\numprint{0.774}} \\ \hline
\nprounddigits{1}\numprint{1.000}K\npnoround & \numprint{0.1} & \nprounddigits{1}\numprint{99.240}K\npnoround & \nprounddigits{1}\numprint{169.147}K\npnoround & \numprint{0.049} & \nprounddigits{1}\numprint{65.230}K\npnoround & \numprint{0.098} & \nprounddigits{1}\numprint{91.652}K\npnoround & \numprint{0.040} & \nprounddigits{1}\numprint{190.040}K\npnoround & \numprint{0.120} & \nprounddigits{1}\numprint{68.706}K\npnoround & \numprint{0.043} & \nprounddigits{1}\numprint{70.571}K\npnoround & \textbf{\numprint{0.024}} \\
\nprounddigits{1}\numprint{1.000}K\npnoround & \numprint{0.2} & \nprounddigits{1}\numprint{1.200585}M\npnoround & \nprounddigits{1}\numprint{2.423957}M\npnoround & \numprint{0.749} & \nprounddigits{1}\numprint{1.241723}M\npnoround & \numprint{1.678} & \nprounddigits{1}\numprint{1.583822}M\npnoround & \numprint{0.518} & \nprounddigits{1}\numprint{2.771750}M\npnoround & \numprint{2.120} & \nprounddigits{1}\numprint{1.335950}M\npnoround & \numprint{0.716} & \nprounddigits{1}\numprint{1.496861}M\npnoround & \textbf{\numprint{0.295}} \\
\nprounddigits{1}\numprint{1.000}K\npnoround & \numprint{0.3} & \nprounddigits{1}\numprint{15.487755}M\npnoround & \nprounddigits{1}\numprint{34.918652}M\npnoround & \numprint{11.211} & \nprounddigits{1}\numprint{21.087409}M\npnoround & \numprint{28.473} & \nprounddigits{1}\numprint{24.994401}M\npnoround & \numprint{7.240} & \nprounddigits{1}\numprint{40.281000}M\npnoround & \numprint{32.961} & \nprounddigits{1}\numprint{23.253962}M\npnoround & \numprint{11.881} & \nprounddigits{1}\numprint{29.796787}M\npnoround & \textbf{\numprint{4.964}} \\ \hline
\nprounddigits{1}\numprint{2.000}K\npnoround & \numprint{0.1} & \nprounddigits{1}\numprint{753.311}K\npnoround & \nprounddigits{1}\numprint{1.404318}M\npnoround & \numprint{0.614} & \nprounddigits{1}\numprint{583.412}K\npnoround & \numprint{1.544} & \nprounddigits{1}\numprint{916.410}K\npnoround & \numprint{0.385} & \nprounddigits{1}\numprint{1.663844}M\npnoround & \numprint{1.447} & \nprounddigits{1}\numprint{610.439}K\npnoround & \numprint{0.589} & \nprounddigits{1}\numprint{633.723}K\npnoround & \textbf{\numprint{0.211}} \\ \hline
\nprounddigits{1}\numprint{3.000}K\npnoround & \numprint{0.1} & \nprounddigits{1}\numprint{2.915372}M\npnoround & \nprounddigits{1}\numprint{5.381197}M\npnoround & \numprint{2.920} & \nprounddigits{1}\numprint{2.303634}M\npnoround & \numprint{8.700} & \nprounddigits{1}\numprint{3.481143}M\npnoround & \numprint{1.614} & \nprounddigits{1}\numprint{6.379637}M\npnoround & \numprint{6.610} & \nprounddigits{1}\numprint{2.389194}M\npnoround & \numprint{3.278} & \nprounddigits{1}\numprint{2.492642}M\npnoround & \textbf{\numprint{1.067}} \\ \hline
\nprounddigits{1}\numprint{10.000}K\npnoround & \numprint{0.001} & \nprounddigits{1}\numprint{50.006}K\npnoround & \nprounddigits{1}\numprint{59.024}K\npnoround & \numprint{0.119} & \nprounddigits{1}\numprint{7.434}K\npnoround & \numprint{0.162} & \nprounddigits{1}\numprint{10.155}K\npnoround & \numprint{1.618} & \nprounddigits{1}\numprint{60.184}K\npnoround & \textbf{\numprint{0.015}} & \nprounddigits{1}\numprint{7.434}K\npnoround & \numprint{0.051} & \nprounddigits{1}\numprint{7.434}K\npnoround & \numprint{0.034} \\
\nprounddigits{1}\numprint{10.000}K\npnoround & \numprint{0.003} & \nprounddigits{1}\numprint{142.630}K\npnoround & \nprounddigits{1}\numprint{155.653}K\npnoround & \numprint{0.219} & \nprounddigits{1}\numprint{13.076}K\npnoround & \numprint{0.478} & \nprounddigits{1}\numprint{17.302}K\npnoround & \numprint{1.659} & \nprounddigits{1}\numprint{159.534}K\npnoround & \textbf{\numprint{0.062}} & \nprounddigits{1}\numprint{13.242}K\npnoround & \numprint{0.145} & \nprounddigits{1}\numprint{13.242}K\npnoround & \numprint{0.082} \\
\nprounddigits{1}\numprint{10.000}K\npnoround & \numprint{0.005} & \nprounddigits{1}\numprint{215.789}K\npnoround & \nprounddigits{1}\numprint{252.616}K\npnoround & \numprint{0.321} & \nprounddigits{1}\numprint{27.713}K\npnoround & \numprint{0.823} & \nprounddigits{1}\numprint{53.187}K\npnoround & \numprint{1.734} & \nprounddigits{1}\numprint{268.407}K\npnoround & \numprint{0.135} & \nprounddigits{1}\numprint{28.619}K\npnoround & \numprint{0.276} & \nprounddigits{1}\numprint{28.616}K\npnoround & \textbf{\numprint{0.133}} \\
\nprounddigits{1}\numprint{10.000}K\npnoround & \numprint{0.010} & \nprounddigits{1}\numprint{349.705}K\npnoround & \nprounddigits{1}\numprint{548.235}K\npnoround & \numprint{0.643} & \nprounddigits{1}\numprint{135.830}K\npnoround & \numprint{2.013} & \nprounddigits{1}\numprint{299.568}K\npnoround & \numprint{2.050} & \nprounddigits{1}\numprint{648.210}K\npnoround & \numprint{0.478} & \nprounddigits{1}\numprint{137.626}K\npnoround & \numprint{0.743} & \nprounddigits{1}\numprint{137.628}K\npnoround & \textbf{\numprint{0.299}} \\
\nprounddigits{1}\numprint{10.000}K\npnoround & \numprint{0.030} & \nprounddigits{1}\numprint{3.746762}M\npnoround & \nprounddigits{1}\numprint{5.201613}M\npnoround & \numprint{5.548} & \nprounddigits{1}\numprint{1.412802}M\npnoround & \numprint{21.601} & \nprounddigits{1}\numprint{1.838754}M\npnoround & \numprint{5.208} & \nprounddigits{1}\numprint{5.577595}M\npnoround & \numprint{8.763} & \nprounddigits{1}\numprint{1.436033}M\npnoround & \numprint{7.617} & \nprounddigits{1}\numprint{1.440393}M\npnoround & \textbf{\numprint{2.316}} \\
\bottomrule
\end{tabular}
\end{table}
\end{landscape}
}

\setlength{\tabcolsep}{1.0ex}

\afterpage{%
\clearpage
\captionsetup{width=\textheight}
\begin{landscape}
\begin{center}
\small
\begin{longtable}{lr@{\hskip 15pt} rr@{\hskip 15pt} rr@{\hskip 15pt} rr@{\hskip 15pt} rr@{\hskip 15pt} rr@{\hskip 15pt} rr}
\caption{Performance of different clique enumeration algorithms over structured graphs. Column header $\mu$ is the number of maximal cliques found. In bold, the best performance for each row. Times are measured in seconds with millisecond precision. A value of `\fail' indicates that the run did not finish within the $6$h time limit. A value of * indicates that the instance could not be run.} \label{table:dimacs} \\

\toprule
\multicolumn{2}{c}{Instance}  & \multicolumn{2}{c}{\alg{Naude}~\cite{naude-2016}} &  \multicolumn{2}{c}{\alg{TomitaBB}}&\multicolumn{2}{c}{\alg{Tomita}~\cite{tomita-2006}}& \multicolumn{2}{c}{\alg{ELS}~\cite{els-2013}} & \multicolumn{2}{c}{\alg{GreedyBBNX}} & \multicolumn{2}{c}{\alg{GreedyBB}} \\
\cmidrule{1-14}
Name & $\mu$ & steps & time & steps & time & steps & time & steps & time & steps & time & steps & time \\
\midrule
\endhead
\multicolumn{14}{r}{{Continued on next page $\rightarrow$}} \\
\endfoot
\bottomrule
\endlastfoot

\texttt{C125.9} & \nprounddigits{1}\numprint{7.451705681}B\npnoround & \nprounddigits{1}\numprint{14.443179474}B\npnoround & \numprint{1338.135} & \nprounddigits{1}\numprint{10.359482442}B\npnoround & \numprint{1868.772} & \nprounddigits{1}\numprint{10.421502187}B\npnoround & \numprint{1588.730} & \nprounddigits{1}\numprint{17.500913643}B\npnoround & \numprint{5614.190} & \nprounddigits{1}\numprint{20.115992693}B\npnoround & \numprint{1262.700} & \nprounddigits{1}\numprint{20.530891116}B\npnoround & \textbf{\numprint{1083.396}} \\
\texttt{MANN\_a9} & \nprounddigits{1}\numprint{590.887}K\npnoround & \nprounddigits{1}\numprint{959.123}K\npnoround & \numprint{0.065} & \nprounddigits{1}\numprint{413.675}K\npnoround & \numprint{0.058} & \nprounddigits{1}\numprint{374.342}K\npnoround & \numprint{0.040} & \nprounddigits{1}\numprint{954.296}K\npnoround & \numprint{0.170} & \nprounddigits{1}\numprint{522.266}K\npnoround & \numprint{0.035} & \nprounddigits{1}\numprint{522.272}K\npnoround & \textbf{\numprint{0.029}} \\ \hline
\texttt{brock200\_1} & \nprounddigits{1}\numprint{449.560064}M\npnoround & \nprounddigits{1}\numprint{934.958134}M\npnoround & \numprint{135.897} & \nprounddigits{1}\numprint{700.437980}M\npnoround & \numprint{232.493} & \nprounddigits{1}\numprint{726.808265}M\npnoround & \numprint{150.180} & \nprounddigits{1}\numprint{1.168729438}B\npnoround & \numprint{505.020} & \nprounddigits{1}\numprint{1.103189008}B\npnoround & \numprint{127.370} & \nprounddigits{1}\numprint{1.792271291}B\npnoround & \textbf{\numprint{116.126}} \\
\texttt{brock200\_2} & \nprounddigits{1}\numprint{431.586}K\npnoround & \nprounddigits{1}\numprint{912.706}K\npnoround & \numprint{0.177} & \nprounddigits{1}\numprint{621.885}K\npnoround & \numprint{0.224} & \nprounddigits{1}\numprint{672.430}K\npnoround & \numprint{0.160} & \nprounddigits{1}\numprint{1.098427}M\npnoround & \numprint{0.550} & \nprounddigits{1}\numprint{757.722}K\npnoround & \numprint{0.129} & \nprounddigits{1}\numprint{1.036765}M\npnoround & \textbf{\numprint{0.084}} \\
\texttt{brock200\_3} & \nprounddigits{1}\numprint{4.595644}M\npnoround & \nprounddigits{1}\numprint{9.941685}M\npnoround & \numprint{1.592} & \nprounddigits{1}\numprint{7.268133}M\npnoround & \numprint{2.554} & \nprounddigits{1}\numprint{7.657120}M\npnoround & \numprint{1.760} & \nprounddigits{1}\numprint{12.163991}M\npnoround & \numprint{5.950} & \nprounddigits{1}\numprint{9.637017}M\npnoround & \numprint{1.392} & \nprounddigits{1}\numprint{15.529450}M\npnoround & \textbf{\numprint{1.107}} \\
\texttt{brock200\_4} & \nprounddigits{1}\numprint{19.645556}M\npnoround & \nprounddigits{1}\numprint{42.035914}M\npnoround & \numprint{6.498} & \nprounddigits{1}\numprint{30.984873}M\npnoround & \numprint{10.759} & \nprounddigits{1}\numprint{32.465381}M\npnoround & \numprint{7.500} & \nprounddigits{1}\numprint{51.803133}M\npnoround & \numprint{24.610} & \nprounddigits{1}\numprint{43.024571}M\npnoround & \numprint{5.823} & \nprounddigits{1}\numprint{70.020160}M\npnoround & \textbf{\numprint{4.848}} \\ \hline
\texttt{c-fat200-5} & \numprint{7} & \numprint{262} & \textbf{$<$\numprint{0.001}} & \numprint{521} & \numprint{0.003} & \numprint{581} & \textbf{$<$\numprint{0.001}} & \numprint{593} & \textbf{$<$\numprint{0.001}} & \numprint{608} & \textbf{$<$\numprint{0.001}} & \nprounddigits{1}\numprint{1.363}K\npnoround & \textbf{$<$\numprint{0.001}} \\
\texttt{c-fat500-5} & \numprint{16} & \numprint{865} & \textbf{$<$\numprint{0.001}} & \nprounddigits{1}\numprint{1.389}K\npnoround & \numprint{0.004} & \nprounddigits{1}\numprint{1.733}K\npnoround & \textbf{$<$\numprint{0.001}} & \nprounddigits{1}\numprint{1.514}K\npnoround & \numprint{0.020} & \nprounddigits{1}\numprint{1.485}K\npnoround & \numprint{0.001} & \nprounddigits{1}\numprint{3.373}K\npnoround & \textbf{$<$\numprint{0.001}} \\
\texttt{c-fat500-10} & \numprint{8} & \numprint{694} & \textbf{$<$\numprint{0.001}} & \nprounddigits{1}\numprint{1.351}K\npnoround & \numprint{0.016} & \nprounddigits{1}\numprint{1.540}K\npnoround & \numprint{0.020} & \nprounddigits{1}\numprint{1.553}K\npnoround & \numprint{0.050} & \nprounddigits{1}\numprint{1.540}K\npnoround & \numprint{0.002} & \nprounddigits{1}\numprint{9.163}K\npnoround & \numprint{0.001} \\ \hline
\texttt{dsjc1000.1} & \nprounddigits{1}\numprint{98.115}K\npnoround & \nprounddigits{1}\numprint{166.807}K\npnoround & \numprint{0.048} & \nprounddigits{1}\numprint{64.107}K\npnoround & \numprint{0.094} & \nprounddigits{1}\numprint{90.018}K\npnoround & \numprint{0.040} & \nprounddigits{1}\numprint{187.457}K\npnoround & \numprint{0.120} & \nprounddigits{1}\numprint{67.528}K\npnoround & \numprint{0.047} & \nprounddigits{1}\numprint{69.290}K\npnoround & \textbf{\numprint{0.025}} \\
\texttt{dsjc1000.5} & \nprounddigits{1}\numprint{10.833119791}B\npnoround & \nprounddigits{1}\numprint{28.549388684}B\npnoround & \numprint{9453.370}& \fail & \fail & \nprounddigits{1}\numprint{22.186328166}B\npnoround & \textbf{\numprint{6422.300}} & \nprounddigits{1}\numprint{2.586921460}B\npnoround & \numprint{25676.120} & \nprounddigits{1}\numprint{24.457159528}B\npnoround & \numprint{14049.509} & \nprounddigits{1}\numprint{47.853975800}B\npnoround & \numprint{6987.053} \\
\texttt{dsjc.125.9} & \nprounddigits{1}\numprint{5.162770941}B\npnoround & \nprounddigits{1}\numprint{10.107502290}B\npnoround & \numprint{934.211} & \nprounddigits{1}\numprint{7.209142333}B\npnoround & \numprint{1262.251} & \nprounddigits{1}\numprint{7.473464654}B\npnoround & \numprint{1126.350} & \nprounddigits{1}\numprint{4.001305588}B\npnoround & \numprint{4076.000} & \nprounddigits{1}\numprint{14.596053290}B\npnoround & \numprint{874.443} & \nprounddigits{1}\numprint{14.984439390}B\npnoround & \textbf{\numprint{755.715}} \\
\texttt{dsjc.250.5} & \nprounddigits{1}\numprint{1.683562}M\npnoround & \nprounddigits{1}\numprint{3.663333}M\npnoround & \numprint{0.626} & \nprounddigits{1}\numprint{2.564436}M\npnoround & \numprint{0.930} & \nprounddigits{1}\numprint{2.740648}M\npnoround & \numprint{0.670} & \nprounddigits{1}\numprint{4.408440}M\npnoround & \numprint{2.450} & \nprounddigits{1}\numprint{3.120898}M\npnoround & \numprint{0.533} & \nprounddigits{1}\numprint{4.575002}M\npnoround & \textbf{\numprint{0.375}} \\
\texttt{dsjc500.1} & \nprounddigits{1}\numprint{14.998}K\npnoround & \nprounddigits{1}\numprint{24.334}K\npnoround & \numprint{0.006} & \nprounddigits{1}\numprint{9.637}K\npnoround & \numprint{0.011} & \nprounddigits{1}\numprint{12.599}K\npnoround & \numprint{0.020} & \nprounddigits{1}\numprint{27.530}K\npnoround & \numprint{0.020} & \nprounddigits{1}\numprint{9.978}K\npnoround & \numprint{0.006} & \nprounddigits{1}\numprint{10.188}K\npnoround & \textbf{\numprint{0.003}} \\
\texttt{dsjc500.5} & \nprounddigits{1}\numprint{102.675431}M\npnoround & \nprounddigits{1}\numprint{245.244718}M\npnoround & \numprint{57.665} & \nprounddigits{1}\numprint{175.564130}M\npnoround & \numprint{102.932} & \nprounddigits{1}\numprint{187.948567}M\npnoround & \numprint{48.980} & \nprounddigits{1}\numprint{287.836619}M\npnoround & \numprint{184.930} & \nprounddigits{1}\numprint{210.431748}M\npnoround & \numprint{57.344} & \nprounddigits{1}\numprint{360.175778}M\npnoround & \textbf{\numprint{39.479}} \\ \hline
\texttt{hamming8-4} & \nprounddigits{1}\numprint{45.215840}M\npnoround & \nprounddigits{1}\numprint{76.584491}M\npnoround & \numprint{8.603} & \nprounddigits{1}\numprint{47.789490}M\npnoround & \numprint{13.374} & \nprounddigits{1}\numprint{49.418122}M\npnoround & \numprint{10.200} & \nprounddigits{1}\numprint{97.522625}M\npnoround & \numprint{41.050} & \nprounddigits{1}\numprint{54.094773}M\npnoround & \numprint{6.095} & \nprounddigits{1}\numprint{57.531373}M\npnoround & \textbf{\numprint{4.632}} \\
\texttt{hamming6-2} & \nprounddigits{1}\numprint{1.281402}M\npnoround & \nprounddigits{1}\numprint{2.503548}M\npnoround & \numprint{0.202} & \nprounddigits{1}\numprint{1.800812}M\npnoround & \numprint{0.224} & \nprounddigits{1}\numprint{1.802427}M\npnoround & \numprint{0.230} & \nprounddigits{1}\numprint{3.131836}M\npnoround & \numprint{0.940} & \nprounddigits{1}\numprint{2.521651}M\npnoround & \numprint{0.156} & \nprounddigits{1}\numprint{2.528347}M\npnoround & \textbf{\numprint{0.125}} \\
\texttt{hamming6-4} & \numprint{464} & \nprounddigits{1}\numprint{1.110}K\npnoround & \textbf{$<$\numprint{0.001}} & \numprint{495} & \textbf{$<$\numprint{0.001}} & \numprint{896} & \textbf{$<$\numprint{0.001}} & \nprounddigits{1}\numprint{1.379}K\npnoround & \textbf{$<$\numprint{0.001}} & \numprint{495} & \textbf{$<$\numprint{0.001}} & \numprint{495} & \textbf{$<$\numprint{0.001}} \\ \hline
\texttt{johnson16-2-4} & \nprounddigits{1}\numprint{2.027025}M\npnoround & \nprounddigits{1}\numprint{12.131448}M\npnoround & \numprint{0.711} & \nprounddigits{1}\numprint{5.863577}M\npnoround & \numprint{1.200} & \nprounddigits{1}\numprint{12.131448}M\npnoround & \numprint{0.670} & \nprounddigits{1}\numprint{14.259609}M\npnoround & \numprint{4.910} & \nprounddigits{1}\numprint{5.863577}M\npnoround & \numprint{0.617} & \nprounddigits{1}\numprint{5.863577}M\npnoround & \textbf{\numprint{0.364}} \\
\texttt{johnson8-2-4} & \numprint{105} & \numprint{380} & \textbf{$<$\numprint{0.001}} & \numprint{205} & \textbf{$<$\numprint{0.001}} & \numprint{380} & \textbf{$<$\numprint{0.001}} & \numprint{505} & \textbf{$<$\numprint{0.001}} & \numprint{205} & \textbf{$<$\numprint{0.001}} & \numprint{205} & \textbf{$<$\numprint{0.001}} \\
\texttt{johnson8-4-4} & \nprounddigits{1}\numprint{114.690}K\npnoround & \nprounddigits{1}\numprint{200.597}K\npnoround & \numprint{0.021} & \nprounddigits{1}\numprint{156.473}K\npnoround & \numprint{0.025} & \nprounddigits{1}\numprint{157.405}K\npnoround & \numprint{0.030} & \nprounddigits{1}\numprint{281.872}K\npnoround & \numprint{0.100} & \nprounddigits{1}\numprint{173.978}K\npnoround & \textbf{\numprint{0.013}} & \nprounddigits{1}\numprint{181.705}K\npnoround & \numprint{0.014} \\ \hline
\texttt{keller4} & \nprounddigits{1}\numprint{10.284321}M\npnoround & \nprounddigits{1}\numprint{13.822807}M\npnoround & \numprint{1.200} & \nprounddigits{1}\numprint{3.958237}M\npnoround & \numprint{1.371} & \nprounddigits{1}\numprint{4.028001}M\npnoround & \numprint{1.250} & \nprounddigits{1}\numprint{14.428810}M\npnoround & \numprint{4.460} & \nprounddigits{1}\numprint{5.403456}M\npnoround & \numprint{0.736} & \nprounddigits{1}\numprint{6.137147}M\npnoround & \textbf{\numprint{0.563}} \\ \hline
\texttt{p\_hat1000-1} & \nprounddigits{1}\numprint{11.079532}M\npnoround & \nprounddigits{1}\numprint{24.199617}M\npnoround & \numprint{7.817} & \nprounddigits{1}\numprint{16.104195}M\npnoround & \numprint{20.825} & \nprounddigits{1}\numprint{17.525171}M\npnoround & \numprint{4.840} & \nprounddigits{1}\numprint{27.775466}M\npnoround & \numprint{17.440} & \nprounddigits{1}\numprint{17.637596}M\npnoround & \numprint{9.629} & \nprounddigits{1}\numprint{22.966003}M\npnoround & \textbf{\numprint{3.801}} \\
\texttt{p\_hat1500-1} & \nprounddigits{1}\numprint{119.749740}M\npnoround & \nprounddigits{1}\numprint{276.173339}M\npnoround & \numprint{103.296} & \nprounddigits{1}\numprint{191.294211}M\npnoround & \numprint{365.288} & \nprounddigits{1}\numprint{204.315904}M\npnoround & \numprint{61.520} & \nprounddigits{1}\numprint{313.025413}M\npnoround & \numprint{218.840} & \nprounddigits{1}\numprint{208.794360}M\npnoround & \numprint{160.645} & \nprounddigits{1}\numprint{289.101936}M\npnoround & \textbf{\numprint{51.242}} \\
\texttt{p\_hat300-1} & \nprounddigits{1}\numprint{58.176}K\npnoround & \nprounddigits{1}\numprint{111.720}K\npnoround & \numprint{0.026} & \nprounddigits{1}\numprint{67.391}K\npnoround & \numprint{0.033} & \nprounddigits{1}\numprint{75.538}K\npnoround & \numprint{0.020} & \nprounddigits{1}\numprint{132.712}K\npnoround & \numprint{0.060} & \nprounddigits{1}\numprint{75.218}K\npnoround & \numprint{0.020} & \nprounddigits{1}\numprint{87.103}K\npnoround & \textbf{\numprint{0.012}} \\
\texttt{p\_hat300-2} & \nprounddigits{1}\numprint{79.917408}M\npnoround & \nprounddigits{1}\numprint{155.300680}M\npnoround & \numprint{22.309} & \nprounddigits{1}\numprint{115.134461}M\npnoround & \numprint{40.961} & \nprounddigits{1}\numprint{116.527175}M\npnoround & \numprint{21.210} & \nprounddigits{1}\numprint{194.880042}M\npnoround & \numprint{68.710} & \nprounddigits{1}\numprint{215.861278}M\npnoround & \numprint{21.752} & \nprounddigits{1}\numprint{240.622053}M\npnoround & \textbf{\numprint{17.167}} \\
\texttt{p\_hat500-1} & \nprounddigits{1}\numprint{548.523}K\npnoround & \nprounddigits{1}\numprint{1.123503}M\npnoround & \numprint{0.254} & \nprounddigits{1}\numprint{724.588}K\npnoround & \numprint{0.482} & \nprounddigits{1}\numprint{794.049}K\npnoround & \numprint{0.210} & \nprounddigits{1}\numprint{1.313583}M\npnoround & \numprint{0.690} & \nprounddigits{1}\numprint{804.554}K\npnoround & \numprint{0.248} & \nprounddigits{1}\numprint{993.989}K\npnoround & \textbf{\numprint{0.148}} \\
\texttt{p\_hat500-2} & \nprounddigits{1}\numprint{59.641233479}B\npnoround & \nprounddigits{1}\numprint{50.607357528}B\npnoround & \numprint{19528.780}& \fail & \fail& \fail & \fail& \fail & \fail& \fail & \fail & \nprounddigits{1}\numprint{217.717864694}B\npnoround & \textbf{\numprint{17219.771}} \\
\texttt{p\_hat700-1} & \nprounddigits{1}\numprint{2.360662}M\npnoround & \nprounddigits{1}\numprint{4.998648}M\npnoround & \numprint{1.319} & \nprounddigits{1}\numprint{3.252353}M\npnoround & \numprint{2.949} & \nprounddigits{1}\numprint{3.579904}M\npnoround & \numprint{0.930} & \nprounddigits{1}\numprint{5.777680}M\npnoround & \numprint{3.350} & \nprounddigits{1}\numprint{3.590444}M\npnoround & \numprint{1.419} & \nprounddigits{1}\numprint{4.558671}M\npnoround & \textbf{\numprint{0.687}} \\ \hline
\texttt{san400\_0.5\_1} & \nprounddigits{1}\numprint{52.937602}M\npnoround & \nprounddigits{1}\numprint{26.688857574}B\npnoround & \numprint{2066.431} & \nprounddigits{1}\numprint{862.834835}M\npnoround & \numprint{2629.124} & \nprounddigits{1}\numprint{866.230052}M\npnoround & \numprint{3097.100} & \nprounddigits{1}\numprint{1.066000910}B\npnoround & \numprint{18209.360} & \nprounddigits{1}\numprint{871.335566}M\npnoround & \numprint{956.527} & \nprounddigits{1}\numprint{874.277749}M\npnoround & \textbf{\numprint{871.392}} \\
\texttt{sanr200\_0.7} & \nprounddigits{1}\numprint{69.575623}M\npnoround & \nprounddigits{1}\numprint{147.874762}M\npnoround & \numprint{23.044} & \nprounddigits{1}\numprint{109.262560}M\npnoround & \numprint{35.342} & \nprounddigits{1}\numprint{114.692218}M\npnoround & \numprint{24.880} & \nprounddigits{1}\numprint{182.890319}M\npnoround & \numprint{84.940} & \nprounddigits{1}\numprint{158.544351}M\npnoround & \numprint{20.114} & \nprounddigits{1}\numprint{262.589802}M\npnoround & \textbf{\numprint{17.589}} \\
\texttt{sanr400\_0.5} & \nprounddigits{1}\numprint{25.120414}M\npnoround & \nprounddigits{1}\numprint{57.876782}M\npnoround & \numprint{12.380} & \nprounddigits{1}\numprint{40.945236}M\npnoround & \numprint{23.753} & \nprounddigits{1}\numprint{43.966570}M\npnoround & \numprint{11.100} & \nprounddigits{1}\numprint{68.448689}M\npnoround & \numprint{41.340} & \nprounddigits{1}\numprint{49.360241}M\npnoround & \numprint{11.806} & \nprounddigits{1}\numprint{78.832933}M\npnoround & \textbf{\numprint{7.717}} \\ \hline
\texttt{4-FullIns\_5} & \nprounddigits{1}\numprint{76.171}K\npnoround & \nprounddigits{1}\numprint{80.179}K\npnoround & \numprint{0.056} & \nprounddigits{1}\numprint{3.720}K\npnoround & \numprint{0.109} & \nprounddigits{1}\numprint{4.470}K\npnoround & \numprint{0.220} & \nprounddigits{1}\numprint{82.213}K\npnoround & \numprint{0.200} & \nprounddigits{1}\numprint{4.809}K\npnoround & \numprint{0.047} & \nprounddigits{1}\numprint{4.840}K\npnoround & \textbf{\numprint{0.024}} \\ 
\texttt{abb313GPIA} & \nprounddigits{1}\numprint{2.590403}M\npnoround & \nprounddigits{1}\numprint{3.773824}M\npnoround & \numprint{0.719} & \nprounddigits{1}\numprint{1.545494}M\npnoround & \numprint{2.033} & \nprounddigits{1}\numprint{1.856827}M\npnoround & \textbf{\numprint{0.320}} & * & * & \nprounddigits{1}\numprint{1.501636}M\npnoround & \numprint{0.557} & \nprounddigits{1}\numprint{1.667929}M\npnoround & \numprint{0.445} \\ \hline
\texttt{flat300\_20\_0} & \nprounddigits{1}\numprint{2.464573}M\npnoround & \nprounddigits{1}\numprint{5.504279}M\npnoround & \numprint{1.038} & \nprounddigits{1}\numprint{3.789259}M\npnoround & \numprint{1.623} & \nprounddigits{1}\numprint{4.123479}M\npnoround & \numprint{1.060} & \nprounddigits{1}\numprint{6.563710}M\npnoround & \numprint{3.810} & \nprounddigits{1}\numprint{4.566236}M\npnoround & \numprint{0.972} & \nprounddigits{1}\numprint{6.838015}M\npnoround & \textbf{\numprint{0.605}} \\
\texttt{flat300\_26\_0} & \nprounddigits{1}\numprint{2.851219}M\npnoround & \nprounddigits{1}\numprint{6.423038}M\npnoround & \numprint{1.205} & \nprounddigits{1}\numprint{4.461008}M\npnoround & \numprint{1.926} & \nprounddigits{1}\numprint{4.834686}M\npnoround & \numprint{1.200} & \nprounddigits{1}\numprint{7.630375}M\npnoround & \numprint{4.480} & \nprounddigits{1}\numprint{5.383175}M\npnoround & \numprint{1.077} & \nprounddigits{1}\numprint{8.270507}M\npnoround & \textbf{\numprint{0.720}} \\
\texttt{flat300\_28\_0} & \nprounddigits{1}\numprint{2.906348}M\npnoround & \nprounddigits{1}\numprint{6.496011}M\npnoround & \numprint{1.206} & \nprounddigits{1}\numprint{4.506846}M\npnoround & \numprint{1.936} & \nprounddigits{1}\numprint{4.885534}M\npnoround & \numprint{1.240} & \nprounddigits{1}\numprint{7.770066}M\npnoround & \numprint{4.550} & \nprounddigits{1}\numprint{5.470801}M\npnoround & \numprint{1.099} & \nprounddigits{1}\numprint{8.318859}M\npnoround & \textbf{\numprint{0.754}} \\
\texttt{flat1000\_50\_0} & \nprounddigits{1}\numprint{7.162568784}B\npnoround & \nprounddigits{1}\numprint{18.928124613}B\npnoround & \numprint{6177.177} & \nprounddigits{1}\numprint{13.575018926}B\npnoround & \numprint{15088.180} & \nprounddigits{1}\numprint{14.686128444}B\npnoround & \textbf{\numprint{4243.590}} & \nprounddigits{1}\numprint{194.301755}M\npnoround & \numprint{18059.100} & \nprounddigits{1}\numprint{16.024144463}B\npnoround & \numprint{11402.193} & \nprounddigits{1}\numprint{31.295128854}B\npnoround & \numprint{4859.020} \\
\texttt{flat1000\_60\_0} & \nprounddigits{1}\numprint{7.706391125}B\npnoround & \nprounddigits{1}\numprint{20.360934724}B\npnoround & \numprint{6645.092} & \nprounddigits{1}\numprint{14.634239387}B\npnoround & \numprint{16286.397} & \nprounddigits{1}\numprint{15.805649715}B\npnoround & \textbf{\numprint{4557.290}} & \nprounddigits{1}\numprint{1.811413574}B\npnoround & \numprint{19268.240} & \nprounddigits{1}\numprint{17.231479634}B\npnoround & \numprint{11151.300} & \nprounddigits{1}\numprint{33.418360676}B\npnoround & \numprint{4888.690} \\
\texttt{flat1000\_76\_0} & \nprounddigits{1}\numprint{8.322676291}B\npnoround & \nprounddigits{1}\numprint{21.967770294}B\npnoround & \numprint{7106.404} & \nprounddigits{1}\numprint{15.824017293}B\npnoround & \numprint{17786.329} & \nprounddigits{1}\numprint{17.054275474}B\npnoround & \textbf{\numprint{4896.330}} & \nprounddigits{1}\numprint{3.659341891}B\npnoround & \numprint{20735.820} & \nprounddigits{1}\numprint{18.651900190}B\npnoround & \numprint{12652.670} & \nprounddigits{1}\numprint{36.256766387}B\npnoround & \numprint{5173.690} \\ \hline
\texttt{qg.order60} & \numprint{120} & \nprounddigits{1}\numprint{73.757}K\npnoround & \numprint{0.052} & \nprounddigits{1}\numprint{110.506}K\npnoround & \numprint{3.106} & \nprounddigits{1}\numprint{80.710}K\npnoround & \numprint{0.320} & \nprounddigits{1}\numprint{4.196996}M\npnoround & \numprint{3.570} & \nprounddigits{1}\numprint{110.506}K\npnoround & \numprint{1.507} & \nprounddigits{1}\numprint{111.676}K\npnoround & \textbf{\numprint{0.027}} \\ \hline
\texttt{r1000.5} & \nprounddigits{1}\numprint{588.533}K\npnoround & \nprounddigits{1}\numprint{1.388184}M\npnoround & \textbf{\numprint{0.954}} & \nprounddigits{1}\numprint{4.308106}M\npnoround & \numprint{15.030} & \nprounddigits{1}\numprint{4.264985}M\npnoround & \numprint{6.410} & \nprounddigits{1}\numprint{5.132604}M\npnoround & \numprint{29.810} & \nprounddigits{1}\numprint{52.099692}M\npnoround & \numprint{6.969} & \nprounddigits{1}\numprint{67.028843}M\npnoround & \numprint{7.349} \\ \hline
\texttt{school1} & \nprounddigits{1}\numprint{247.543890}M\npnoround & \nprounddigits{1}\numprint{348.816708}M\npnoround & \numprint{33.710} & \nprounddigits{1}\numprint{110.100662}M\npnoround & \numprint{54.634} & \nprounddigits{1}\numprint{115.125521}M\npnoround & \textbf{\numprint{20.630}} & \nprounddigits{1}\numprint{350.803817}M\npnoround & \numprint{70.420} & \nprounddigits{1}\numprint{238.274633}M\npnoround & \numprint{28.156} & \nprounddigits{1}\numprint{286.437596}M\npnoround & \numprint{26.230} \\
\texttt{school1\_nsh} & \nprounddigits{1}\numprint{33.118973}M\npnoround & \nprounddigits{1}\numprint{48.126110}M\npnoround & \numprint{4.230} & \nprounddigits{1}\numprint{15.078093}M\npnoround & \numprint{6.588} & \nprounddigits{1}\numprint{16.738480}M\npnoround & \textbf{\numprint{2.630}} & \nprounddigits{1}\numprint{48.822947}M\npnoround & \numprint{10.310} & \nprounddigits{1}\numprint{39.177358}M\npnoround & \numprint{4.027} & \nprounddigits{1}\numprint{47.920229}M\npnoround & \numprint{3.810} \\ \hline
\texttt{wap03a} & \nprounddigits{1}\numprint{84.137}K\npnoround & \nprounddigits{1}\numprint{355.002}K\npnoround & \numprint{0.439} & \nprounddigits{1}\numprint{521.421}K\npnoround & \numprint{6.349} & \nprounddigits{1}\numprint{455.800}K\npnoround & \textbf{\numprint{0.610}} & \nprounddigits{1}\numprint{4.196996}M\npnoround & \numprint{3.510} & \nprounddigits{1}\numprint{693.677}K\npnoround & \numprint{2.989} & \nprounddigits{1}\numprint{3.346848}M\npnoround & \numprint{0.803} \\
\texttt{wap04a} & \nprounddigits{1}\numprint{84.700}K\npnoround & \nprounddigits{1}\numprint{364.049}K\npnoround & \numprint{0.454} & \nprounddigits{1}\numprint{513.087}K\npnoround & \numprint{6.960} & \nprounddigits{1}\numprint{453.101}K\npnoround & \textbf{\numprint{0.690}} & \nprounddigits{1}\numprint{4.196996}M\npnoround & \numprint{3.530} & \nprounddigits{1}\numprint{673.534}K\npnoround & \numprint{3.366} & \nprounddigits{1}\numprint{3.195455}M\npnoround & \numprint{0.830} \\
\end{longtable}
\end{center}
\end{landscape}
}

\end{document}